%
%
%
\magnification=\magstephalf
\hyphenation{brem-sstrahlung}
\def\nohyphen{\hyphenpenalty=10000}

\hsize=6.5truein 
\vsize=9.15truein
\hoffset=-0.1truein
\voffset=-.15truein 

\baselineskip=24truept plus0truept minus0truept

\parindent=3em
\overfullrule=0pt
\headline={\ifodd\pageno\rightheadline\else\rightheadline\fi}
\def\numbers{\def\rightheadline{\hfil\tenrm\folio\hfil}} 
\def\nonumbers{\def\rightheadline{\hfil}} 
\nonumbers 
\nopagenumbers 
\tolerance=800 
\hbadness=5000
\def\refs{\baselineskip=15truept plus0truept minus0truept
    \leftskip=2.7em\parindent=-2.7em\parskip=8pt plus0truept minus0truept}
\def\sig{\sigma_{_{\rm T}}}

\def\gapprox{\lower.4ex\hbox{$\;\buildrel >\over{\scriptstyle\sim}\;$}}
\def\lapprox{\lower.4ex\hbox{$\;\buildrel <\over{\scriptstyle\sim}\;$}}
\def\Msun{M_\odot}
\def\part#1#2{{\partial#1\over\partial#2}}
\def\frac#1#2{{#1 \over #2}}

\def\gapprox{\lower.4ex\hbox{$\;\buildrel >\over{\scriptstyle\sim}\;$}}
\def\lapprox{\lower.4ex\hbox{$\;\buildrel <\over{\scriptstyle\sim}\;$}}
\def\pt{\ifmmode \tilde{p} \else $\tilde{p}$~ \fi}

\def\lsim{\lower.5ex\hbox{$\;\buildrel < \over \sim \;$}}
\def\gsim{\lower.5ex\hbox{$\; \buildrel> \over \sim \;$}}

\def\cm{\ifmmode {\rm cm}^{-1} \else cm$^{-1}$ \fi}
\def\s{\ifmmode {\rm s}^{-1} \else s$^{-1}$ \fi}
\def\cc{\ifmmode {\rm cm}^{-3} \else cm$^{-3}$ \fi}
\def\cs{\ifmmode {\rm cm}^{-2} \else cm$^{-2}$ \fi}

\def\g{\ifmmode \gamma \else $\gamma$\fi}
\def\G{\ifmmode \Gamma \else $\Gamma$\fi}
\def\kms{\ifmmode {\rm km\ s}^{-1} \else km s$^{-1}$\fi}

\def\gad{\ifmmode \Gamma_{ad} \else $\Gamma_{ad}$ \fi}
\def\g{\ifmmode \gamma \else $\gamma$\fi}
\def\G{\ifmmode \Gamma \else $\Gamma$\fi}

\def\Mout{\dot M_{\rm out}}
\def\Min{\dot M_{\rm in}}
\def\Rout{R_{\rm out}}
\def\Rin{R_{\rm in}}

\nohyphen
\phantom{XXXX}
\vskip1.5truein
\centerline{\bf FORMATION OF RELATIVISTIC OUTFLOWS IN}
\centerline{\bf SHEARING BLACK HOLE ACCRETION CORONAE}

\bigskip

\centerline{\bf Prasad Subramanian$^1$, Peter A. Becker$^2$}

\bigskip

\centerline{Center for Earth Observing and Space Research,}
\centerline{Institute for Computational Sciences and Informatics,}
\centerline{George Mason University, Fairfax, VA 22030-4444}

\bigskip

\centerline{$^1$ \it also \rm Code 7660, Naval Research Laboratory,}
\centerline{Washington, DC 20375,
psubrama@mahanadi.nrl.navy.mil}

\bigskip

\centerline{$^2$ \it also \rm Department of Physics and Astronomy,}
\centerline{George Mason University, Fairfax, VA 22030-4444,
pbecker@gmu.edu}

\bigskip

\centerline{\bf Demosthenes Kazanas}
\bigskip
\centerline{Laboratory for High-Energy Astrophysics,}
\centerline{NASA/Goddard Space Flight Center, Greenbelt, MD 20771,}
\centerline{kazanas@milkyway.gsfc.nasa.gov}

\bigskip
\bigskip
\bigskip
\bigskip
\bigskip
\bigskip
\bigskip
\bigskip

\centerline{Received \underbar{\phantom{XXXXXXXXX}};
accepted\underbar{\phantom{XXXXXXXXX}}}

\bigskip
\bigskip

\centerline{accepted for publication in the Astrophysical Journal}

\vfil
\eject

\numbers
\pageno=2
\centerline{ABSTRACT}

We examine the possibility that the relativistic jets observed in
many active galactic nuclei may be powered by the Fermi acceleration
of protons in a tenuous corona above a two-temperature accretion disk.
In this picture the acceleration arises as a consequence of the shearing
motion of the magnetic field in the corona, which is anchored in the
underlying Keplerian disk. The protons in the corona have a power-law
distribution because the density there is too low for proton-proton
collisions to thermalize the energy supplied via Fermi acceleration.
The same shear acceleration mechanism also operates in the disk itself,
however, there the density is high enough for thermalization to occur and
consequently the disk protons have a Maxwellian distribution.

Particle acceleration in the corona leads to the development of a
pressure-driven wind that passes through a critical point and subsequently
transforms into a relativistic jet at large distances from the black hole.
We combine the critical conditions for the wind with the structure equations
for the disk and the corona to obtain a coupled disk/corona/wind model. Using
the coupled model we compute the asymptotic Lorentz factor $\Gamma_\infty$ of
the jet as a function of the cylindrical starting radius at the base of the
outflow, in the corona. Our results suggest that $\Gamma_\infty \lapprox 10$,
which is consistent with observations of superluminal motion in
blazars. We show that collisions between the jet and broad-line
emission clouds can produce high-energy radiation with a luminosity
sufficient to power the $\gamma$-rays observed from blazars.

Subject headings: radiation mechanisms: non-thermal,
accretion, accretion disks, acceleration of particles,
gamma rays: theory 

\vfil
\eject

\bigskip
\centerline{\bf 1. INTRODUCTION}
\bigskip

Jets are among the most ubiquitous of astrophysical phenomenon, and are
associated with objects ranging from protostars and binary systems in
our galaxy to supermassive black holes in the centers of Active Galactic
Nuclei (AGNs). Large asymptotic Lorentz factors are implied by
observations of superluminal motion in many blazars (Dermer \&
Schlickeiser 1992; Vermeulen \& Cohen 1994), and the {\sl EGRET}
instrument on board the {\sl Compton Gamma-Ray Observatory} ({\sl
CGRO}) has also detected intense $\gamma$-ray flares from many of these
sources (Wehrle et al. 1998). The mechanisms responsible for producing
the observed jets are still poorly understood, but collimated outflows
in general appear to be associated with objects that derive their
luminosity from the accretion of matter onto a gravitating object (e.g.,
Dermer \& Schlickeiser 1992). The asymptotic jet velocity observed far
from the central object is probably related to the depth of the
gravitational potential at the radius of jet formation. The observations
of relativistic jets with asymptotic Lorentz factors $\Gamma_\infty
\lapprox 10$ suggest an origin relatively close to the black hole
horizon, implying the existence of extreme conditions at the base of the
outflow. Despite the overall consensus that AGNs presumably contain
accretion disks, little, if any, connection has been established between
the observed jets and the underlying disks. Our goal in this paper is to
explore the possibility that the energy responsible for powering the
jets is transferred to the jet particles via a second-order Fermi
acceleration mechanism driven by the shear of the accretion disk.

There has been a significant amount of recent interest in interpreting
the observed high-energy emission from blazars as arising from hadronic
jets (e.g., Mannheim 1993; Dar \& Laor 1997), and a variety of different
mechanisms have been proposed to channel the required energy from the
protons into the $\gamma$-rays. A complete theoretical understanding of
the emission mechanisms operating in these sources is going to be
particularly important for the interpretation of the wealth of data
expected from the upcoming {\sl GLAST} observations of AGNs. It seems
clear that while the X-ray and soft $\gamma$-ray ($\lapprox 10\,$MeV)
components of typical AGN spectra can be successfully attributed to
leptonic models, the GeV and TeV emission are better explained by
hadronic models, involving either proton-induced cascades (Mannheim
1993) or collisions between the hadronic jet and target protons located
in the broad-line emission region (Dar \& Laor 1997). One of the reasons
hadronic models are better suited to explain the GeV and TeV emission is
that hadrons do not suffer from the severe radiative losses associated
with electrons, and therefore hadrons do not require large in situ
reacceleration rates. Furthermore, GeV energies emerge as natural scales
in hadronic processes such as the strong proton-proton interaction.
However, the origin of the energetic hadrons in the jets is ambiguous.
Most models utilizing protons beams make the ad hoc assumption that the
protons are acceleration by shocks running up and down the jet. While
shock acceleration might very well be taking place, the connection
between the jet and the underlying accretion disk which surrounds the
central object still remains unclear.

In principle, radiation pressure provides another means for accelerating
the gas, but this mechanism alone probably cannot power a highly
relativistic outflow regardless of whether it is optically thin or
thick. Under optically thin conditions, the aberration of the background
photons will cause them to approach from the head-on direction in the
frame of the outflowing gas, impeding rather than accelerating a
relativistic outflow (Phinney 1982). Conversely, under optically thick
conditions the photon energy density at the base of the flow must exceed
the rest mass energy density by at least one order of magnitude in order
to achieve an asymptotic Lorentz factor $\Gamma_\infty \sim 10$ (cf.
M\'esz\'aros \& Rees 1992). This probably cannot occur in the accretion
disk, where we expect rough equipartition between the matter and the
radiation. 

Magnetocentrifugal acceleration may provide another mechanism for
producing relativistic jets (Blandford \& Payne 1982; Contopoulos \&
Lovelace 1994; Contopoulos 1995). In this scenario, energy associated
with the rotation of a cold accretion disk is transferred via magnetic
stresses to the tenuous gas in an overlying corona, resulting in the
acceleration of the gas to relativistic speeds in a magnetically
confined jet (Li, Chiueh, \& Begelman 1992). Most theoretical treatments
of magnetocentrifugal acceleration make drastic simplifications, such as
assuming self-similar dependences of the flow parameters on the radial
distance. These treatments require the poloidal magnetic field at the
disk surface to be inclined at a sufficiently large angle away from the
$z$-axis in order for the acceleration mechanism to be effective. The
simulations of Ustyugova et al. (1995) suggest that the asymptotic flow
velocities attained via the magnetocentrifugal acceleration mechanism
are only mildly relativistic at best. Furthermore, the stability of
these magnetically confined flows is uncertain (Begelman 1998), and it
is unclear how the existence of the relativistic outflow is related to
the values of the macroscopic physical parameters in the underlying
accretion flow.

Hydrodynamical acceleration driven by the pressure of the hot gas itself
provides a third mechanism for the production of relativistic outflows.
This has been proposed by Mastichiadis \& Kazanas (1993) and developed
in greater detail by Contopoulos \& Kazanas (1995). In these models the
required heating of the gas is accomplished in situ via shocks or
turbulence (Protheroe \& Kazanas 1983; Kazanas \& Ellison 1986), and the
outflow may also receive additional power from the decay of relativistic
neutrons throughout the jet. Within the context of this general
scenario, the thickness of the accretion disk divided by the neutron
flight distance determines whether highly relativistic outflows are
possible in a given situation (Contopoulos \& Kazanas 1995). However, in
general, these ``distributed power'' models lack a detailed connection
with the physics of the underlying accretion disk.

In the present paper we examine a new type of disk/jet connection
by considering a simple model for the production of a
relativistic outflow based on the hydrodynamical expansion of
a gas which is heated via a Fermi acceleration mechanism. This
is dynamically similar to the distributed-power models except
that the heating takes place at the base of the outflow only,
where it is possible to establish a direct connection between
the required proton acceleration and the properties of the
accretion disk. The outflowing gas must have positive energy in
order to escape from the black hole, so that the resulting wind
is super-virial (e.g., Blandford \& Begelman 1999).
Romanova et al. (1998) performed numerical simulations of
the dynamics of magnetic loop configurations in the coronae of accretion
disks. They found that the loops open up due to differential rotation in the
disk, leading to transient outflows driven by pressure from
toroidal magnetic fields.

In the model considered here the energy
is also supplied via differential rotation in the disk, but
the specific mechanism involved is second-order Fermi acceleration
occurring in the tenuous corona due to the motion of magnetic
field lines anchored in the disk. Microphysically, the Fermi acceleration results
from collisions between seed (thermal) protons and magnetic scattering
centers (``kinks'') whose motion is driven by the Keplerian shear flow
of the underlying disk. In this model, the magnetic field produces a
viscous torque on the disk, and the dissipated azimuthal kinetic energy
is transferred to the corona primarily by the Poynting flux of the
shearing field rather than by the thermal protons diffusing up from the
disk. We show that this process results in a relativistic
power-law tail in the coronal proton distribution, with enough pressure
to drive the particles out of the corona as a transonic electron-proton
wind. Beyond the sonic (critical) point, the wind transitions into a
relativistic outflow with an asymptotic bulk Lorentz factor $\Gamma_\infty
\sim 3 - 10$. The scenario of protons interacting with the tangled magnetic
field has been considered by Subramanian, Becker, \& Kafatos (1996, hereafter
Paper I) as one of the mechanisms responsible for providing the
viscosity in the accretion disk. Hence the same physical process which
accomplishes the transfer of angular momentum in the disk also produces
the acceleration of the relativistic protons in the corona. 

The remainder of the paper is organized as follows.
In \S~2 we present a detailed analysis of the correspondence between
the viscosity mechanism introduced in Paper I and the second-order Fermi
acceleration mechanism considered in this paper. After we have established the 
equivalence of the two scenarios, we focus in \S~3 on solution
of the steady-state transport equation describing the energy
distribution of the protons accelerated in the corona. In \S~4
we develop the formalism required to self-consistently link the
corona with the underlying disk, resulting
in a coupled disk-corona model. The details of the wind/jet
formation process are considered in \S~5, and in \S~6 we discuss
the model self-consistency requirements. Computational results
obtained using the fully coupled disk/corona/wind model are
presented in \S~7, and in \S~8 we conclude with a discussion of
our results and the associated implications for $\gamma$-ray
observations of blazars.

\bigskip
\centerline{\bf 2. FERMI ACCELERATION AND VISCOSITY}
\bigskip

We begin by investigating the connection between the viscosity mechanism
discussed in Paper I and the Fermi acceleration scenario considered in
this paper by comparing the heating rates associated with the two
processes. As in Paper 1, we will focus on hot, two-temperature
accretion disks with ion (proton) temperatures $T_i \sim 10^{12}\,$K and
electron temperatures $T_e \sim 10^9\,$K. We are mainly concerned here
with the energetic consequences of collisions between hot protons and
kinks in the magnetic field that are participating in the shear flow of
the Keplerian disk. In order to focus on the direct effects of the
magnetic shear, we assume that the kinks are ``cold,'' meaning that they
have no stochastic motion. The validity of this assumption will be
investigated later when we consider the consequences of replacing the
cold kinks with propagating, stochastic MHD waves. Although our specific
application involves the acceleration of protons in a corona overlying
the disk, our discussion of the shear acceleration mechanism will remain
general at this point.

\medskip
\centerline{2.1. \it Fermi Acceleration in Shear Flows}
\medskip

A qualitative argument for the second-order nature of the Fermi
acceleration mechanism operating in a shear flow can be constructed as
follows. Consider a proton originating in the (stationary) middle layer
in Figure~1 and experiencing a subsequent collison with a scattering
center (cold magnetic field kink) located in Quadrant 2. Since this is
an approaching collision, the proton will gain energy. Conversely, the
corresponding collision in Quadrant 1 is an overtaking one, and therefore
the proton will lose energy in this case. Following this chain of
logic for the other two quadrants, we conclude that to first order in
the relative shear velocity between successive scattering centers,
$\Delta v$, there is no mean gain or loss of energy for the incident
proton. However, the approaching collisions take place on a shorter
timescale than the overtaking ones, and therefore, to second order in
$\Delta v$, acceleration dominates over deceleration.

The theoretical basis for second-order Fermi acceleration due to
collisions with scattering centers embedded in a shear flow has been
examined in the context of cosmic ray energization by Earl, Jokipii, \&
Morfill (1988) and by Webb, Jokipii, \& Morfill (1994), and the application to accretion
flows has been discussed by Katz (1991). Typically, second-order Fermi
acceleration occurs when particles interact with randomly-moving
scattering centers, whereas the interaction of particles with
systematically-moving scattering centers (e.g. in a converging flow)
usually results in first-order Fermi acceleration. In our situation, the
scattering centers (kinks in the tangled magnetic field) are embedded in
a systematic (Keplerian) flow, and are cold (no stochastic motion).
Nonetheless, as discussed above, the interaction results in a mean
fractional energy gain per scattering $\Delta \epsilon/\epsilon \propto
\Delta v^2$, where $\epsilon$ is the proton energy. The acceleration
mechanism considered here is therefore hybrid in nature since it is a
{\it second-order} process operating in a {\it systematic} background flow.

When the scattering centers are contained in a Keplerian shear flow,
the mean fractional energy gain per scattering is given by
$$
\frac{\Delta \epsilon}{\epsilon} \sim \left(\Delta v_{\phi}
\over c\right)^2 = \left( {\tilde\lambda \over c}
{dv_{\phi} \over dR} \right)^2 \,, \eqno(2.1)
$$
where $\tilde{\lambda}$ is the mean free path for collisions between
protons and magnetic scattering centers, $c$ is the speed of light,
$v_\phi$ denotes the Keplerian orbital velocity, and $\Delta v_{\phi}
\equiv \tilde{\lambda} \, (dv_{\phi}/dR)$ gives the characteristic
relative shear velocity between successive scattering centers. In this
type of situation, we can model the diffusion of the protons in energy
space using a simple transport equation of the form
$$
\frac{\partial f}{\partial t} = - \frac{1}{\epsilon^2}
\frac{\partial}{\partial \epsilon} \left( - \epsilon^2 {\cal D}
\frac{\partial f}{\partial \epsilon} \right) \,, \eqno(2.2)
$$
where ${\cal D}$ is the energy diffusion coefficient and the distribution
function $f$ is related to the ion number density $N$ and energy density
$U$ by
$$
N = \int_{m_p \, c^2}^{\infty} \epsilon^2 \, f \, d \epsilon
\, \, \, \sim \, \, {\rm cm^{-3}} \,, \eqno(2.3)
$$
$$ 
U = \int_{m_p \, c^2}^{\infty} \epsilon^3 \, f \, d \epsilon
\, \, \, \sim \, \, {\rm ergs \ \ cm^{-3}} \,, \eqno(2.4)
$$
with $m_p$ denoting the proton mass. Note that equation~(2.2) considers
only diffusion in energy space, and ignores spatial transport. We will
have occasion later in the paper to replace the lower bound of integrals
like those in equations~(2.3) and (2.4) with zero, because the
mathematical structure of the relevant equations will allow for the
diffusion of particles to negligibly small energies. We will not,
however, be making a serious error by adopting a lower bound of zero in
such situations, because we will be dealing with relativistic proton
distributions containing very few particles with energies close to $m_p
c^2$.

\vfil
\eject

\medskip
\centerline{2.2. \it Energy Diffusion Coefficient}
\medskip

We can quantify the energy diffusion coefficient ${\cal D}$ introduced
in equation~(2.2) by relating it to the fractional energy change per
scattering given by equation~(2.1). Using equation~(A2), we express
the mean energization rate due to shear acceleration for protons with
energy $\epsilon$ as
$$
\langle \dot\epsilon_{\rm shear} \rangle = \frac{1}{\epsilon^2}
\frac{d}{d \epsilon} \left( \epsilon^2 {\cal D} \right)
= 4 \, \epsilon \, D \,, \eqno(2.5)
$$
where we have adopted the form for the energy diffusion coefficient
$$
{\cal D}(\epsilon) = D \, \epsilon^2 \ \ \sim \ \ 
{\rm ergs^2 \ s^{-1}} \eqno(2.6)
$$
with $D=\,$constant, which is appropriate for cases involving
an energy-independent magnetic scattering cross section. Note that
$\langle \dot\epsilon_{\rm shear} \rangle \propto \epsilon$, as is
typical of Fermi processes. We can write another expression for the
acceleration rate based directly on equation~(2.1), which yields
$$
\frac{\Delta \epsilon}{\Delta t} = \epsilon \,
\left(\tilde\lambda \over c\right) \left(\frac{dv_{\phi}}{dR}\right)^2
\,, \eqno(2.7)
$$
where $\Delta t = \tilde\lambda / c$ is the mean free time for
collisions between protons and magnetic scattering centers. Equating the
acceleration rates given by equations~(2.5) and (2.7), we find that
$$
D = \frac{1}{4} \left(\tilde\lambda \over c\right)
\left(\frac{dv_{\phi}}{dR}\right)^2 \ \ \sim \ \ 
{\rm s^{-1}}\,, \eqno(2.8)
$$
which establishes the nature of the diffusion
coefficient governing the stochastic transport of the Fermi
accelerated protons through the energy space.
In the case of the relativistic particles,
the heating rate due to Fermi acceleration can be obtained by
averaging equation~(2.5) over the particle energy and
multiplying by the proton number density, which yields
$$
\frac{dU}{dt} = 4 \, \langle\epsilon\rangle \, D \, N
= 4 \, D \, U \ \ \sim \ \ {\rm ergs \ \ cm^{-3}
\ \ s^{-1}} \,, \eqno(2.9)
$$
where $\langle\epsilon\rangle \equiv U/N$ is the mean proton energy.

\medskip
\centerline{2.3. \it Viscous Heating vs. Nonthermal Particle Acceleration}
\medskip

Although we have derived our formal results for Fermi acceleration
under the assumption that the protons are relativistic, we can
calculate the corresponding energy diffusion coefficient
for the thermal protons in the shear
flow by replacing the relativistic particle velocity $c$ in equation~(2.8)
with the typical thermal velocity $v_{\rm rms} = \sqrt{3 k T_i / m_i}
= \lambda_{ii} / t_{ii}$, where $\lambda_{ii}$ is the mean free path
for proton-proton collisions and
$$
t_{ii} = 11.4 \, {T_i^{3/2} \over N \ln\Lambda} \ \ \sim
\ \ {\rm s} \eqno(2.10)
$$
is the associated mean free time for Coulomb logarithm $\ln\Lambda$
(see Paper I). The result obtained for the thermal diffusion coefficient
is
$$
D_{\rm th} = 2.85 \, {\tilde\lambda \over \lambda_{ii}}
\, {T_i^{3/2} \over N \ln\Lambda} \,
\left(\frac{dv_{\phi}}{dR}\right)^2 \,. \eqno(2.11)
$$
The corresponding heating rate associated with the Fermi acceleration
of the thermal protons can be obtained by writing in analogy to equation~(2.9)
$$
\frac{dU_{\rm th}}{dt} = 4 \, D_{\rm th} \, U_{\rm th}
= 2.36 \times 10^{-15} \, {\tilde\lambda \over \lambda_{ii}}
\, {T_i^{5/2} \over \ln\Lambda} \,
\left(\frac{dv_{\phi}}{dR}\right)^2 \,, \eqno(2.12)
$$
where $U_{\rm th} = (3/2) N k T_i$ is the energy density of the thermal
protons. This result for the Fermi heating rate of the thermal protons
is comparable to the viscous dissipation rate computed using the usual
formula $\eta_{\rm hyb}(dv_{\phi}/dR)^2$ (Landau \& Lifshitz 1987),
where $\eta_{\rm hyb}$ is the ``hybrid'' viscosity coefficient derived
in Paper~I for the same physical situation considered here, but using a
standard approach based upon an explicit calculation of the momentum
flux in the shear flow. In the hybrid scenario, the angular momentum
is carried by the protons, but the collisions occur predominantly between
the protons and the magnetic field kinks. We have therefore established
that the shear-driven Fermi acceleration experienced by the thermal
protons is equivalent to conventional viscous heating.

The Fermi acceleration mechanism pumps energy into all of the protons
residing in either the disk or the corona, since both of these regions
are threaded by the shearing magnetic field. Although each region
experiences the same shear, the shape of the resulting steady-state
proton distribution will also depend on the degree to which the energy
supplied by Fermi acceleration is thermalized via proton-proton
collisions. In the disk, the proton number density is high enough for
collisional thermalization to occur, resulting in a Maxwellian
distribution for the disk protons (Katz 1991). Conversely, due to the
low density in the corona, we do not expect thermalization to occur very
effectively there, and consequently the coronal protons are likely to
have a nonthermal distribution. In \S~3 we support this conclusion by
carefully considering the losses occurring in the corona.

\bigskip
\centerline{\bf 3. KINETIC EQUATION}
\bigskip

In the general picture considered here, thermal seed
protons from the accretion disk diffuse upwards into the tenuous corona,
where they undergo second-order Fermi acceleration due to collisions
with kinks in the local magnetic field. We assume that the field lines
in the corona are anchored in the underlying disk, and that the lines
(and the kinks) therefore participate in the same shearing motion as the
disk itself. Under suitable conditions, the protons accelerated in the
corona possess enough energy to drive a transonic electron-proton wind
that can escape to infinity with a sizable asymptotic Lorentz factor.
The shear-driven second-order Fermi acceleration mechanism will tend
to produce a nonthermal proton energy distribution in the corona,
although the shape of the distribution is also affected by
particle interactions which can thermalize the relativistic protons.

\medskip
\centerline{3.1. \it Collisional Losses}
\medskip

The collisional losses experienced by protons in the corona occur
primarily via interactions with other protons mediated by either
the Coulomb force or the strong force. The
ratio of the loss timescales for these two processes can
be expressed as a function of the proton energy $\epsilon =
\gamma m_p c^2$ using (e.g., Dermer, Miller, \& Li 1996)
$$
\frac{t_{\rm Coul}}{t_{\rm pp}} = 1225 \, \frac{\sigma_{\rm pp}}
{\sig} \, \frac{\gamma - 1}
{\beta^2 \, \ln\Lambda} \, \biggl(3.8 \, \theta_{\rm pl}^{3/2}
+ \beta^3 \biggr ) \,, \eqno(3.1)
$$
where $t_{\rm Coul}$ is the timescale for Coulomb collisions, $t_{\rm
pp}$ is the timescale for strong interactions, $\sigma_{\rm pp}=3
\times 10^{-26}\,{\rm cm}^2$ is the strong interaction cross section,
$\sig$ is the Thomson scattering cross section, $\beta \equiv
(1 - \gamma^{-2})^{1/2}$, and $\theta_{\rm pl} \equiv k T_i / m_e c^2$,
with $m_e$ denoting the electron mass. Setting $T_i = 10^{12}\,$K and
$\ln\Lambda = 25$, we find that losses due to strong interactions
dominate over proton-proton Coulomb losses for all of the energetic
protons, whether in the disk or the corona. This is a direct consequence
of the virial temperature of the hot disk. Conversely, in
relatively cool plasmas, Coulomb losses dominate (Dermer, Miller, \& Li
1996).

The energy loss rate due to strong proton-proton interactions
can be written as
$$
\dot{\epsilon}_{\rm loss} = - \nu_{\rm loss} \, \epsilon
\,, \ \ \ \ \ \ \nu_{\rm loss} = \sigma_{\rm pp} \, c \, N \,.
\eqno(3.2)
$$
The direct dependence of the strong interaction loss rate upon the
proton number density $N$ implies that the collisional losses will
be more severe in the dense disk midplane than in the relatively
tenuous corona, and we therefore assume that the disk protons have
a Maxwellian distribution with temperature $T_i$. The importance
of collisional losses in the corona depends on the density there,
and in \S~6 we estimate an upper limit for the accretion rate
below which the loss timescale due to strong proton-proton interactions
in the corona exceeds the shear acceleration timescale. We are justified
in neglecting proton losses in our calculations so long as we confine
our attention to accretion rates that are well below this upper limit,
which greatly exceeds the Eddington accretion rate.

In addition to the energy losses associated with proton-proton
interactions, the relativistic protons in the corona also experience
losses due to Coulomb coupling with the attendant electrons
(which cool readily via synchrotron and inverse-Compton emission)
and in principle this can limit the Lorentz factor of any resulting jet
(Phinney 1982). However, so long as the electron and proton
number densities are equal, proton-electron Coulomb losses are
less important than proton-proton Coulomb losses (Schmidt
1966), which are negligible compared with the losses due to strong
proton-proton
interactions according to equation~(3.1). This conclusion must be
modified if copious electron-positron pair production takes place
in the corona, since this process can significantly enhance the
proton-electron cooling rate by increasing the number of electrons (and
positrons) per proton. We address this issue in \S~6 by demonstrating
that the optical thickness of the corona to photon-photon pair
production is much less than unity, implying that pair production can be
safely ignored. In principle, the protons also lose energy via {\it
direct} synchrotron and inverse-Compton emission, but these processes
are unimportant in the proton energy range of interest here.

\vfil
\eject

\medskip
\centerline{3.2. \it Particle Distribution in the Corona}
\medskip

The transport equation governing the energy distribution of the relativistic
protons in the corona can be obtained by incorporating a source term
and an escape term into equation~(2.2) and substituting for the energy
diffusion coefficient $\cal D$ using equation~(2.6), which yields
$$
{\partial f \over \partial t} =
{D_1 \over \epsilon^2} \, {\partial \over \partial \epsilon} \,
\left( \epsilon^4 \, {\partial f \over \partial \epsilon}
\right) - \, {f \over t_1} \,
+ \frac{\dot N_1 \, \delta (\epsilon - \epsilon_i)}
{\epsilon_i^2} \,, \eqno(3.3)
$$
where the subscript ``1'' will be used to denote parameters
associated with the corona, and $t_1$ represents the mean time protons
spend in the corona before escaping into the wind. The final term
in equation~(3.3) describes the injection of monoenergetic protons
with energy $\epsilon_i$ at a rate per unit volume equal to
$\dot N_1$. Physically, the ``injection'' process corresponds to
the diffusion of seed protons from the underlying disk up into the
corona. Note that a Maxwellian cannot be recovered from equation~(3.3)
because it does not contain a collisonal loss term; our neglect of
collisional losses is reasonable for sufficiently low accretion rates,
as discussed in \S~6. We can relate the escape timescale $t_1$ to the
coherence length in the corona $\tilde\lambda_1$ using
$$
t_1 \equiv {h \over v_1}
= {h^2 \over c \, \tilde\lambda_1} \,, \eqno(3.4)
$$
where $h$ is the vertical thickness of the corona and the diffusion
velocity for protons escaping from the corona into the wind is given by
$$
v_1 \equiv {c \, \tilde \lambda_1 \over h}
\sim {\kappa_1 \over h} \,, \eqno(3.5)
$$
with
$$
\kappa_1 \equiv \frac{1}{3} \, c \, \tilde\lambda_1 \eqno(3.6)
$$
denoting the associated spatial diffusion coefficient (e.g., Reif 1965).
We shall assume throughout that the corona has no vertical structure.

Before solving equation~(3.3) for the proton energy distribution
in the corona, it is interesting to calculate the mean energy of
the protons in the corona based on fundamental properties of the
Fermi acceleration process. As a first step, consider the time
evolution of the mean energy of a {\it single} proton injected into the
corona with energy $\epsilon_i$ at time $t=0$. By integrating
equation~(2.5) we find that the mean energy of this proton (while
it remains in the corona) varies as
$$
\bar \epsilon (t) = \epsilon_i \, e^{4 D_1 t} \,.
\eqno(\rm 3.7)
$$
In the absence of losses, the mean energy of the proton therefore
increases without bound until the proton escapes from the corona
by diffusing out into the wind. Next we make use of the observation
that the mean energy of the protons in the corona is equal to the
mean energy of the protons escaping from the corona, which is a
consequence of the fact that the escape timescale $t_1$ is
independent of the proton energy. We can therefore compute
the mean energy of the protons in the corona by calculating
the average energy of the single proton at the time that it
escapes from the corona. This yields
$$
\langle \epsilon \rangle = \int_0^\infty \bar \epsilon
(t) \, e^{-t/t_1} \, {dt \over t_1} \,, \eqno(\rm 3.8)
$$
where the last factor expresses the probability that the proton
will remain in the corona until time $t$ and then escape
during the subsequent time interval $dt$. Substituting
for $\bar \epsilon (t)$ in equation~(3.8) using
equation~(3.7) and integrating yields
$$
\langle \epsilon \rangle = {\epsilon_i \over 1 - y} \,,
\eqno(\rm 3.9)
$$
where we have defined the $y$-parameter for the Fermi process
as
$$
y \equiv 4 \, D_1 \, t_1
=  {\langle \dot\epsilon_{\rm shear} \rangle \over \epsilon} \,
t_1 \,, \eqno(\rm 3.10)
$$
and made use of equation~(2.5) to arrive at the final result.

In the analogous case of photon Comptonization, the $y$-parameter
measures the mean fractional energy gain experienced by soft photons
scattering in a medium of hot electrons, and $y$ must exceed unity for
significant distortion of the input spectrum to occur (e.g., Rybicki \&
Lightman 1979). However, for the shear-driven Fermi acceleration process
treated here, the mean proton energy $\langle \epsilon \rangle$ {\it
diverges} as $y \to 1$. From a physical point of view, the divergence
of $\langle \epsilon \rangle$ is due to the fact that an infinitesimal
number of protons remain in the corona long enough to gain an infinite
amount of energy, which leads to a logarithmic divergence in the total
energy density as $y \to 1$. Hence in our model $y$ cannot exceed (or
even equal) unity. The apparent contradiction is due to the fact that
electron recoil losses are included in the Comptonization model, whereas
the corresponding loss mechanisms for the protons (direct synchrotron and
inverse-Compton emission) are unimportant in the context of the particle
transport model considered here. Note that the Fermi $y$-parameter is
independent of the mean free path $\tilde\lambda_1$, since it can be
rewritten as
$$
y = \left({h \over c} \,
{dv_\phi \over dR} \right)^2 \,, \eqno(\rm 3.11)
$$
where we have used equations~(2.8) and (3.4). The detailed numerical
results presented in \S~7 indicate that $y \sim 1$ throughout most of the
corona, and therefore we conclude based on equation~(3.9) that the average
energy of the protons in the corona is much higher than that in the disk.

In a steady-state ($\partial/\partial t \rightarrow 0$), the transport
equation~(3.3) can be written in the form
$$
D_1 \, \epsilon^{2} \, \frac{\partial^2 f}
{\partial
\epsilon^2} + 4 \, D_1 \, \epsilon \,
 \frac{\partial f}{\partial \epsilon} - {f \over t_1}
=- \frac{\dot N_1 \, \delta (\epsilon - \epsilon_i)}
{\epsilon_i^2} \,. \eqno(3.12)
$$
The homogeneous equation obtained when
$\epsilon \ne \epsilon_i$ admits the power-law solutions
$$
G_A (\epsilon, \epsilon_i) \equiv
{\dot N_1 \over \epsilon_i^3} \,
{(\epsilon/\epsilon_i)^{m_A} \over
D_1 \, (m_A - m_B)} \ , \ \ \ \ 
G_B (\epsilon, \epsilon_i) \equiv
{\dot N_1 \over \epsilon_i^3} \,
{(\epsilon/\epsilon_i)^{m_B} \over
D_1 \, (m_A - m_B)} \ , \eqno(3.13)
$$
where
$$
m_A \equiv - \frac{3}{2} + \sqrt{ \frac{9}{4} + \frac{4}{y}} \ ,
\ \ \ \ \ 
m_B \equiv - \frac{3}{2} - \sqrt{ \frac{9}{4} + \frac{4}{y}} \ .
\eqno(3.14)
$$
Under the restriction $y < 1$ required in order to obtain a finite value
for $\langle \epsilon \rangle$ using equation~(3.9), we find that $m_A >
1$ and $m_B < -4$. Second-order Fermi acceleration is a stochastic process,
and therefore some of the protons lose energy and some gain energy.
It follows from the constraints on $m_A$ and $m_B$ that
$G_A$ describes the distribution for
$\epsilon \le \epsilon_i$ and $G_B$ describes the distribution for
$\epsilon \ge \epsilon_i$. The result obtained for the Green's function
is therefore
$$
G(\epsilon, \, \epsilon_i) = \cases{
G_A (\epsilon, \epsilon_i) \ , &$\epsilon \le \epsilon_i$\ , \cr
\phantom{G_A(\epsilon, \epsilon_i) \ ,} &\phantom{
$\epsilon \le \epsilon_i$\ ,} \cr
G_B(\epsilon, \epsilon_i) \ , &$\epsilon \ge \epsilon_i$\ . \cr
} \eqno(3.15)
$$

The total particle number density associated with the Green's
function is given by
$$
N_{\rm G} \equiv \int_0^\infty \epsilon^2 \, G(\epsilon, \, \epsilon_i)
\, d \epsilon = \dot N_1 \, t_1 \,, \eqno(3.16)
$$
in agreement with the result obtained by operating on equation~(3.12)
with $\int_0^\infty \epsilon^2 \, d \epsilon$. We can also integrate
the Green's function to obtain the associated energy density,
$$
U_{\rm G} \equiv \int_0^\infty \epsilon^3 \, G(\epsilon, \, \epsilon_i)
\, d \epsilon = {\dot N_1 \, \epsilon_i \, t_1
\over 1 - 4 \, D_1 \, t_1} \,, \eqno(3.17)
$$
which can be verified by operating on equation~(3.12)
with $\int_0^\infty \epsilon^3 \, d \epsilon$.
We remind the reader of the arguments made immediately following
equation~(2.4), concerning the lower bound of integration
in equations~(3.16) and (3.17). The mean energy of the protons
is therefore
$$
\langle \epsilon \rangle = {U_{\rm G} \over N_{\rm G}}
= {\epsilon_i \over 1 - 4 \, D_1 \, t_1} \,, \eqno(3.18)
$$
in agreement with equation~(3.9).

The Green's function given by equation~(3.15) represents the response
of equation~(3.12) to the diffusion of monoenergetic seed particles
from the disk into the corona. The particles injected into the
corona via the diffusion process appear at a rate per unit volume
equal to $\dot N_1$. Although we have assumed that the seed particles
are monoenergetic, in reality the protons diffusing into the corona
have a thermal distribution of energies corresponding to the disk
temperature $T_i$. However, this does not cause any difficulties
because equation~(3.12) is linear, and therefore the proton distribution
resulting from the diffusion of the thermal disk protons into the
corona can be obtained by convolving the Green's function with the
Maxwellian source distribution. We can accomplish this formally by
writing
$$
f(\epsilon) = \int_0^\infty {G(\epsilon, \, \epsilon_i)
\over \dot N_1} \, S(\epsilon_i,T_i) \, d \epsilon_i \,, \eqno(3.19)
$$
where $S(\epsilon_i,T_i) \, d \epsilon_i$ gives the number of protons
in the energy range between $\epsilon_i$ and $\epsilon_i + d \epsilon_i$
appearing in the corona per unit volume per unit time due to the
Maxwellian source. In this paper it is our intention to focus mainly
on the dynamics of the disk/corona/wind system, and we therefore defer
a detailed calculation of the coronal proton distribution to a subsequent
paper. However, we expect that the energy distribution of the protons
in the corona will be described by the power law behavior $f \propto
\epsilon^{m_B}$ at high energies, with the power-law index $m_B$
given by equation~(3.14). At lower energies (approaching the mean
energy of the disk protons), the coronal proton distribution will
retain a Maxwellian form.

\bigskip
\centerline{\bf 4. DISK-CORONA STRUCTURE}
\bigskip

In order to obtain a coupled disk/corona/wind model, we must understand
how the disk structure influences the diffusion of the thermal seed
protons into the corona. In the scenario envisioned here, the protons
diffusing into the corona have a Maxwellian energy distribution at the
local disk temperature $T_i(R)$, where $R$ is the cylindrical radius
at the point of interest in the corona. The mean energy of the escaping
protons is equal to $(3/2)k T_i$, assuming that the disk protons are
nonrelativistic. Since the mean energy of the escaping protons is equal
to the mean energy of the protons remaining in the disk, the escape of
protons from the disk has no effect on the disk temperature. However,
this process will affect the pressure and density distributions in the
disk, and the loss of mass will cause the accretion rate to decrease
with decreasing radius. We must therefore perform a self-consistent
calculation of the structures of the disk and the corona.

\medskip
\centerline{4.1. \it Disk Structure}
\medskip

Motivated by earlier studies of two-temperature flows (e.g., Shapiro, Lightman,
\& Eardley 1976;
Eilek \& Kafatos 1983; Paper 1), we adopt the $\alpha$-disk
model (see Frank, King, \& Raine 1992 for a review). In particular, we
assume that the ions are virially hot ($T_i \sim 10^{12}\,$K), while the
electrons are able to cool effectively via inverse-Compton and
synchrotron emission and therefore have a much lower temperature ($T_e
\sim 10^9\,$K). For simplicity, we shall assume that the disk possesses
no vertical structure, and that it is composed of fully-ionized
hydrogen, with internal energy density $U_0$, pressure $P_0=(\gamma_0-1)
\, U_0$, proton number density $N_0$, and mass density $\rho_0=m_p \,
N_0$. The quantities $U_0$, $P_0$, $N_0$, $\rho_0$, and the disk
half-thickness $H$ are all functions of the cylindrical radius $R$.
The nonrelativistic value of the adiabatic index ($\gamma_0 = 5/3$)
will be used throughout to describe the thermodynamic properties of
the gas in the disk. The geometry of the disk-corona system is indicated
in Figure~2. According to the virial theorem, the internal energy of
the disk protons is equal to half their gravitational potential energy,
so that
$$
U_0 = \frac{1}{2} \, \frac{N_0 \, m_p \, c^2}{R_*-2} \,, \eqno(4.1)
$$
where $R_* \equiv R \, c^2/GM$ is the dimensionless radius. In writing
equation~(4.1), we have approximated the effects of general relativity
by using the pseudo-Newtonian prescription for the gravitational
potential (Paczynski \& Wiita 1980).

The Keplerian azimuthal velocity corresponding to the pseudo-Newtonian
potential is given by (Paczynski \& Wiita 1980)
$$
{v_{\phi} \over c} = {\sqrt{R_*} \over R_*-2} \,, \eqno(4.2)
$$
and we can use equation~(4.1) to write the classical adiabatic sound
speed in the disk as
$$
s_0 \equiv \sqrt{\gamma_0 \, P_0 \over \rho_0}
=c \ \sqrt{\gamma_0 (\gamma_0 - 1) \over 2 \, (R_* - 2)}
\,. \eqno(4.3)
$$
By combining equations~(4.2) and (4.3) we can express the azimuthal
Mach number as
$$
{\cal M}_\phi \equiv {v_\phi \over s_0} =
\sqrt{{2 \over \gamma_0 (\gamma_0-1)} {R_* \over R_*-2}} \,.
\eqno(4.4)
$$
The usual scaling relations for the disk half-thickness $H$ and the
radial drift velocity $v_R$ remain valid for a flow governed by the
pseudo-Newtonian potential, so that we have (see Frank, King, \& Raine
1992)
$$
H \sim {\cal{M}_{\phi}}^{-1} R \,, \eqno(4.5)
$$
$$
v_R \sim \alpha {\cal{M}_{\phi}}^{-1} s_0 \,. \eqno(4.6)
$$
Combining these relations with equations (4.3) and (4.4), we find that
the radial velocity and the height are given by
$$
{v_R \over c} = \alpha \, \frac{\gamma_0 (\gamma_0 - 1)}
{2\,\sqrt{R_*}} \,, \eqno(4.7)
$$
$$
{H \over R} = \sqrt{{\gamma_0 (\gamma_0 - 1) \over 2} \,
{R_* - 2 \over R_*}} \,. \eqno(4.8)
$$
Note that $H / R \ll 1$ throughout the flow, and that $v_R$ is defined
to be positive for infall.

\medskip
\centerline{4.2. \it Disk-Corona Connection}
\medskip

Second-order Fermi acceleration of the relativistic protons takes
place in the corona as a result of multiple collisions with
magnetic scattering centers (kinks) dragged along by field lines
anchored in the shearing disk. The corona has thickness $h$, energy
density $U_1$, pressure $P_1 = (\gamma_1-1) \, U_1$, proton number
density $N_1$, and mass density $\rho_1$, where $\gamma_1 = 4/3$
since the protons are relativistic. In keeping with our approximate
model for the disk (which has no vertical structure),
we assume for simplicity that the corona also has no
vertical structure. Standard disk/corona models
require vertical hydrostatic equilibrium, and this leads to the conclusion
that the gas pressure in the disk exceeds that in the corona.
However, in the model considered here, the gas in the corona has
positive specific energy as a consequence of Fermi acceleration,
and therefore the protons in the corona are not bound to the black hole.
Hence we cannot use hydrostatic equilibrium to determine the pressure
in the corona. As an alternative prescription, we invoke a pressure
balance between the disk and the corona ($P_0 = P_1$) so that no net
force exists between these regions. This assumption is physically
reasonable since any imbalance would be quickly removed by sound
waves propagating across the disk/corona interface.

It follows from the statement of pressure equilibrium and the values
of $\gamma_0$ and $\gamma_1$ that
$$
U_1 = 2 \, U_0 \,. \eqno(4.9)
$$
In the absence of a pressure gradient, the transport of
seed particles from the disk to the corona occurs primarily via spatial
diffusion, and the efficiency of the diffusion process depends on
the degree to which the field is tangled in the disk. Likewise, the
escape of relativistic protons from the corona into the base of the wind
can also be treated as a diffusive process, with an efficiency that
depends on the degree to which the field is tangled in the corona.
The efficiency of diffusion is determined primarily by the coherence
length of the magnetic field, $\tilde\lambda$. In most simulations of
fully-developed MHD turbulence driven by the magnetic shearing
instability, $\tilde\lambda$ scales as a fixed fraction
of the size of the computational ``box,'' which
corresponds to $H$ in the disk or to $h$
in the corona (Brandenburg et al. 1995; Hawley, Gammie, \& Balbus
1995; Matsumoto
\& Tajima 1995). In their linear stability analysis, Matsumoto \&
Tajima (1995) find that the fastest growing modes have a 
$\tilde\lambda \sim 0.1 \, L$, where $L$ is the characteristic
size of the computational box. We therefore model the magnetic
field by introducing the two parameters
$$
\xi_0 \equiv {\tilde\lambda_0 \over H} \,, \ \ \ \ \ \ 
\xi_1 \equiv {\tilde\lambda_1 \over h} \,, \eqno(4.10)
$$
which describe the degree of tangling in the disk and the corona,
respectively. Note that for a diffusive prescription to apply, clearly
$\xi_0$ and $\xi_1$ cannot exceed unity.

We can use equations~(3.5) and (4.10) to define the diffusion
velocities in the disk and the corona, respectively, as
$$
v_0 \equiv \xi_0 \, c \,, \ \ \ \ \ \ \ \ v_1 \equiv
\xi_1 \, c \,. \eqno(4.11)
$$
Vertical hydrostatic equilibrium in the disk is maintained so long
as $v_0 \ll v_R$, which is seen to be satisfied in our results.
Conservation of the proton flux in a column connecting the disk to the
corona implies that the proton number densities in the disk ($N_0$) and
in the corona ($N_1$) are related via the one-dimensional continuity
equation
$$
N_1 \, v_1 = N_0 \, v_0 \,. \eqno(4.12)
$$
It follows from equations~(4.11) and (4.12) that the densities in the
two regions are related by
$$
{N_1 \over N_0} = {\xi_0 \over \xi_1} \,. \eqno(4.13)
$$
We generally expect to find that $N_1 / N_0 < 1$ since the particles
populating the corona escape from the disk via diffusion. In this case
our assumption of pressure equilibrium ($P_0 = P_1$) implies that the
energy per particle in the corona exceeds that in the disk, which is
crucial for the formation of a relativistic wind (e.g., Contopoulos \&
Kazanas 1995) as discussed in \S~5.

The acceleration of the protons in the corona obviously requires a
substantial source of energy, and in our model we implicitly assume
that the energy is supplied to the corona via the Poynting flux of
the shearing magnetic field, which drains kinetic energy from the
Keplerian accretion disk via viscous dissipation. Although it is
not our goal here to develop a detailed model for the magnetic field
in the disk and the corona, we can use our assumption of disk/corona
pressure equilibrium along with our results concerning Fermi acceleration
to constrain the thickness of the corona, $h$, based upon energy
considerations. Let us assume for the moment that the protons diffusing
into the corona are monoenergetic, so that we can use equations~(3.16)
and (3.17) to write
$$
N_1 = N_{\rm G} = \dot N_1 \, t_1 \,, \ \ \ \ \ \ \ \ \ 
U_1 = U_{\rm G} = {\dot N_1 \, \epsilon_i \, t_1 \over
1 - 4 \, D_1 \, t_1}\,, \eqno(4.14)
$$
where $\dot N_1$ denotes the rate per unit volume at which
particles with energy $\epsilon_i$ diffuse into the corona and
the escape timescale for the protons in the corona is given by
$$
t_1 \equiv {h \over v_1} = {h \over c \, \xi_1} \,. \eqno(4.15)
$$
Combining equations~(4.13) and (4.14) with the pressure equilibrium
condition expressed by equation~(4.9) yields
$$
2 = {U_1 \over U_0} = {\xi_0 \over \xi_1}\,{1 \over
1 - y} \,, \eqno(4.16)
$$
where $y = 4 \, D_1 \, t_1$ according to equation~(3.10) and we have
used the fact that $U_0 = N_0 \, \epsilon_i$.
Although we have assumed monoenergetic injection in deriving
equation~(4.16), the result is valid for protons injected
with an arbitrary energy spectrum. We can use equation~(2.8)
to obtain another expression for $D_1$ under the assumption
of Keplerian flow. Incorporating the pseudo-Newtonian potential
(eq.~[4.2]) and assuming that the velocity differential between the
footpoints of the tangled magnetic field is transmitted unchanged
from the disk to the corona, we obtain
$$
D_1 = {\tilde\lambda_1 c \over R_g^2} \,
{(R_* + 2)^2 \over 16 \, R_* \, (R_* - 2)^4} \,, \eqno(4.17)
$$
where $R_g \equiv GM/c^2$. Eliminating $D_1$ between equations~(4.16)
and (4.17) and setting
$\tilde\lambda_1 = h \xi_1$ and $t_1 = h/(\xi_1 c)$, we find
that the extent of the acceleration region $h$ is given by
$$
{h \over R_g} = 2 \, \sqrt{R_*} \, \frac{(R_*-2)^2}{R_*+2}
\left(1 - {\xi_0 \over 2 \, \xi_1} \right)^{1/2} \,. \eqno(4.18)
$$

\medskip
\centerline{4.3. \it Variation of the Accretion Rate}
\medskip

The escape of protons from the accretion flow into the wind via
the corona will cause the disk accretion rate $\dot M$ to decrease
with decreasing radius. Since the structure of the disk (and
therefore the conditions in the corona) will in turn be influenced by
the variation of the accretion rate, we must determine $\dot M$
in a self-consistent manner. The continuity equation relating a
differential change in the disk accretion rate to a differential
change in the wind mass loss rate is given by
$$
d\dot m = - \, {1 \over 2} \, d \dot M \,, \eqno(4.19)
$$
where $\dot m$ denotes the mass loss rate into one of the two winds
emanating from the disk-corona system. The factor of 1/2 appears because
we have only taken into account the mass lost into one of the winds.
The flux of protons leaving the disk is equal to $N_0 v_0$, which
must equal $N_1 v_1$ by virtue of equation~(4.12). It follows that
the continuity equation for the disk can be written as
$$
d \dot M = 2 \cdot m_p \, N_1 \, v_1
\cdot 2 \pi \, R \, dR \,. \eqno(4.20)
$$

The accretion
rate can also be expressed in terms of the disk parameters as
$$
\dot M = 2 \pi R \cdot 2 H \cdot m_p \, N_0 \, v_R \,, \eqno(4.21)
$$
which can be combined with equations~(4.7), (4.8), and (4.20) to obtain
a differential equation for the variation of the accretion rate,
$$
\frac{d \, {\rm ln} \, \dot{M}}{dR} = \left( \frac{9}{5} \right)^{3/2}
\, 
\frac{\xi_{0}}{\alpha} \frac{1}{\sqrt{R_*-2}}
\, {1 \over R_g} \,, \eqno(4.22)
$$
with solution
$$
{\dot M(R) \over \Mout} \, = \exp
\left\{2 \, \left(9 \over 5\right)^{3/2} \,
{\xi_0 \over \alpha} \left[\left({R \over R_g} - 2\right)^{1/2}
- \left({\Rout \over R_g} - 2\right)^{1/2} \right] \right\} \,, \eqno(4.23)
$$
where $\Rout$ is the largest radius at which shear acceleration
in the corona is strong enough to expel particles into the wind
(i.e., the wind ``turn-on'' radius), and $\Mout \equiv \dot M(\Rout)$.
With $\dot M(R)$ determined, we can calculate $\dot m(R)$ using
equation~(4.19), which yields
$$
\dot m(R) = {\Mout - \dot M(R) \over 2} \,. \eqno(4.24)
$$
Since no mass is lost into the wind at radii beyond $\Rout$, we can
interpret $\Mout$ as the accretion rate supplied to the disk at a
large distance from the central object. Equation~(4.23) expresses
the subsequent decrease in the disk accretion rate $\dot M$ as a
result of the expulsion of mass from radius $\Rout$ inwards. Note
that in deriving this result we have assumed that $\xi_0$ is
independent of radius and we have set $\gamma_0 = 5/3$. We consider
the procedure for calculating the wind turn-on radius $\Rout$
in \S~6 where specific applications are made.

\bigskip
\centerline{\bf 5. RELATIVISTIC WIND FORMATION}
\bigskip

Protons energized via second-order Fermi acceleration in the shearing
corona will escape via spatial diffusion, which tends to oppose the
density gradient and will therefore transport particles preferentially
in the ``upward'' direction, leading to the formation of a rotating,
transonic wind. In the picture developed here, the diffusion velocity
$v_1$ for the protons escaping from the corona is set equal to the
vertical component of the flow velocity at the base of the wind, so that
the corona and the wind merge smoothly. The assumption of a smooth
wind-corona merger is physically reasonable since there is no ``hard''
interface between the two regions. The outflowing proton
distribution is expected to remain isotropic in the comoving frame due
to interactions with magnetic kinks, lending validity to a fluid
description of the wind plasma. As the gas moves away from the central
object, it expands and cools adiabatically, since protons are not
subject to strong radiative losses. Through the expansion process,
the internal energy of the protons is gradually converted into kinetic
energy of the outflow. In order for the gas to escape to infinity,
the internal energy per unit mass in the corona must be super-virial,
so that the corona is not bound to the black hole. This is equivalent
to the statement that the {\it total} energy per unit mass in the corona
must be positive (Blandford \& Begelman 1999).

The detailed streamline shape which the outflow follows depends on
the nature of the collimation mechanism, which is far from clear.
Collimation of the flow could occur via magnetic hoop stresses if
the jet is enveloped by toroidal magnetic fields in a magnetic
``cocoon'' (Blandford \& Payne 1982; Contopoulos \& Lovelace 1994;
Ustyugova et al. 1995). The cocoon models rely upon magneto-centrifugal
acceleration and assume the existence of a large-scale magnetic field
that guides the jet. The non-magnetic models of Daly \& Marscher (1988)
display an oscillating cross section which depends upon the details
of the balance between the gas pressure in the jet and the pressure of
the external medium. Given the large uncertainty in the details of the
outflow cross section, we present a theory in which the streamline shape
is left arbitrary. As in Chakrabarti (1985), the inherently
three-dimensional problem is reduced to a two-dimensional problem by
assuming azimuthal symmetry. The two-dimensional problem, in turn, is
reduced to a one-dimensional problem by prescribing the shape of the
``streamtube,'' which is the locus of streamlines that the gas parcels
follow from a given starting radius in the corona. Although the magnetic
field may play an important role in determining the streamtube shape, we
assume here that the magnetic pressure is negligible compared with the
pressure of the protons.
The streamtube shape is left arbitrary in our theory, although in
our specific examples we present results associated with conical
outflow (e.g., Mannheim 1993; Blandford \& Konigl 1979), with a
variety of opening angles.

\medskip
\centerline{5.1. \it Wind Structure Equations}
\medskip

While viscous dissipation plays a crucial role in determining the
structure of the underlying accretion disk, we do not expect shear
stresses to be important in the wind because the density of the gas
drops very rapidly during the expansion. We therefore assume that the
rotating wind emanating from the corona is inviscid, so that the
specific angular momentum $\lambda$ is conserved along the streamtubes.
We further assume that the base of the wind (the corona) corotates
with the Keplerian disk, and that this corotation is enforced by the
magnetic field lines that thread the corona and are anchored in the
disk. We therefore set $\lambda$ equal to its Keplerian value at
the base of a streamtube as part of our condition for smoothly merging
the corona with the wind, and consequently $\lambda$ will depend on the
starting radius in the corona.

The streamtube shape is defined by postulating the cylindrical radius of
the streamline $r$ to be a specified function $f$ of the height above
the disk midplane $z$ and the cylindrical starting radius at the base
of the flow $R$,
$$
r \equiv f\, (R, z) \,. \eqno(5.1)
$$
Defining $z_1$ to be the height at which the corona connects with
the wind, it follows that
$$
R = f\, (R, z_1) \,. \eqno(5.2)
$$
The components of the flow velocity in the
wind in the $\hat z$, $\hat r$, and $\hat\phi$
directions will be denoted by $v$, $u$, and
$w$, respectively.

Introducing the convenient definitions
$$
f_{z} \equiv \left(  \frac{\partial f}{\partial z} \right)_{R}, \, \,
f_{R} \equiv \left(  \frac{\partial f}{\partial R} \right)_{z}, \, \,
f_{z \, R} \equiv \frac{\partial^{2} \, f}{\partial z \partial
R}\ \ , \, \, f_{z \, z} \equiv \frac{\partial^{2} \, f}{\partial z^{2}}
\ \ , \eqno(5.3)
$$
we note that
$$
u = v \, f_{z} \,, \eqno(5.4)
$$
and therefore the bulk Lorentz factor of the flow is given by
$$
\Gamma = {c \over \sqrt{c^2 - w^2 - (1 + f_z^2) v^2}} \,. \eqno(5.5)
$$
Utilizing the relativistically correct definition of the angular
momentum per unit mass,
$$
\lambda = \Gamma \, r \, w \,, \eqno(5.6)
$$
and assuming that the wind is inviscid ($\lambda=$constant),
we can rewrite our expression for $\Gamma$ as
$$
\Gamma^2 = {c^2 + \lambda^2 / f^2 \over c^2 - v^2 \, (1 + f_z^2)} \,.
\eqno(5.7)
$$
In our calculations, we set the specific angular momentum equal
to its Keplerian value in the pseudo-Newtonian potential (Paczynski
\& Wiita 1980),
$$
\lambda = {G M \over c} \, \frac{R_*^{3/2}}{R_* - 2}  \,. \eqno(5.8)
$$

If we restrict our attention to cylindrical annuli along the flow,
then the conservation equation for the proton number density in the
wind ($N$) can be written as (e.g., Mihalas \& Mihalas 1984)
$$
\Gamma \, r \, N \, v \, dr = {\rm constant} \,. \eqno(5.9)
$$
This expression, which decribes the conservation of $z$-directed particle
flux, can be rewritten using equations~(5.1) and (5.3) as
$$
N \, v = N_1 \, v_1 \, \frac{\Gamma_1}{\Gamma} \,
{R \over f} \, {1 \over f_R} \,, \eqno(5.10)
$$
where, as before, the subscript ``1'' denotes quantities measured in
the corona, which is also the base of the wind. In the case of purely
radial outflow, with $f(R, z) = R \, z / z_1\,$, it can be verified
that the number conservation equation reduces to the standard mass
conservation equation for relativistic Bondi flow (Subramanian 1997).

Since the outflow is powered by the pressure of the relativistic
protons, we set the adiabatic index in the wind equal to the
relativistic value used in the corona, $\gamma_1 = 4/3$. As the
gas expands, the protons cool adiabatically because losses due to
Coulomb coupling with the electrons or the direct emission of radiation
are negligible. The classical adiabatic sound speed $s$ in the wind
can therefore be expressed as a function of $z$ and $v$ using
$$
s \equiv \left(\gamma_1 \, P \over N \, m_p\right)^{1/2}
= s_1 \, \left({\Gamma_1 \over \Gamma}
\, {v_1 \over v} \, {R \over f} \, {1 \over f_R} \right)
^{(\gamma_1 - 1)/2} \,, \eqno(5.11)
$$
where the final result follows from equation~(5.10) in combination
with the adiabatic law $P \propto N^{\gamma_1 - 1}$. We remind the
reader that the classical adiabatic sound speed $s$ is related to
the relativistic sound speed $a$ by (Weinberg 1972)
$$
a^2 = {(\gamma_1 - 1) \, c^2 \, s^2 \over
(\gamma_1 - 1) \, c^2 + s^2} \,. \eqno(5.12)
$$
When the flow is adiabatic as assumed here, the wind obeys the
relativistic Bernoulli equation (Mihalas \& Mihalas 1984; Contopoulos
\& Kazanas 1995)
$$
\left(1 + {s^2 \over c^2}{1 \over \gamma_1 - 1} \right) \, \Gamma
- {R_g \over \sqrt{f^2 + z^2} - 2 \, R_g} \equiv B(v,z)
= {\rm constant} \,, \eqno(5.13)
$$
where the second term on the left-hand side represents the
pseudo-Newtonian potential and the quantity $\sqrt{f^2 + z^2}$
denotes the spherical radius. In adiabatic outflows, the Bernoulli
function $B(v,z)$ is conserved
along streamtubes, although it varies as a function of the starting radius
in the corona. The Bernoulli constant is equal to the dimensionless
energy per unit mass at the base of the wind, and it must therefore be
positive if the flow is to escape to infinity (Blandford \& Begelman
1999). Once the streamtube function $f$ has been specified, the Lorentz
factor $\Gamma$ appearing in equation~(5.13) can be expressed as a
function of $z$ and $v$ using equation~(5.7) and therefore
equation~(5.13) can be used to determine the variation of the vertical
velocity $v$ as a function of the height $z$. Note that in order to
solve for the velocity we must first determine the value of the
Bernoulli constant corresponding to critical flow.

\medskip
\centerline{5.2. \it Critical Point Conditions}
\medskip

In order for the outflow to reach an infinite distance from the central
mass, the velocity must increase monotonically as a function of radius.
This requires the wind to pass through a sonic point, which is a
critical point of the governing differential equation as was first noted
by Bondi (1952) in his analysis of spherically symmetric,
nonrelativistic flows. We can derive the appropriate set of critical
point conditions by differentiating $B(v,z)$ with respect to $z$ and
$v$, which yields an equation of the form
$$
\frac{d \, v}{d \, z} = \frac{\partial B/ \partial z}{\partial
B/ \partial v} \,. \eqno(5.14)
$$
Critical points occur where the numerator and the denominator of
equation~(5.15) vanish simultaneously. Setting $\partial B/
\partial v = 0$ yields
$$
v_c^2 \left[1 + f_z^2 (R, z_c) \right]
= a_c^2
= {c^2 \, (\gamma_1 - 1) \, s_c^2 \over (\gamma_1 - 1) \, c^2
+ s_c^2} \,, \eqno(5.15)
$$
where the subscript ``$c$'' denotes quantities measured at the
critical point, $z = z_c$. This equation implies that the velocity
in the $(r, \, z)$ plane is equal to the relativistic sound speed
at $z = z_c$, which is analogous to the condition $v_c = s_c$
encountered in nonrelativistic, nonrotating flows. Setting
$\partial B/ \partial z = 0$ at the critical point yields
the additional constraint
$$
\left(1 + {2 - \gamma_1 \over \gamma_1 - 1} \, {s^2 \over c^2}\right)
{\Gamma f_z f^2 \over f^2 + \lambda^2 / c^2}
\left(\Gamma^2 v^2 f_{zz} - {\lambda^2 \over f^3}\right)
- \Gamma \, s^2 \left({f_z \over f} + {f_{zR} \over f_R}\right)
$$ 
$$
+ {(f \, f_z + z) \, R_g \, c^2 \over \sqrt{f^2 + z^2} \,
(\sqrt{f^2 + z^2} - 2 \, R_g)^2}
\bigg |_{z = z_c} = 0 \,. \eqno(5.16)
$$

Since the Bernoulli function is conserved along streamtubes, we can
obtain another useful relation by writing
$$
B (v_1, \, z_1) = B (v_c, \, z_c) \,, \eqno(5.17)
$$
which is a statement of energy conservation. The sound speed at
the base of the flow, $s_1$, can be expressed in terms of quantities
measured at the critical point using equation~(5.11), which yields
$$
s_1 = s \, \left({\Gamma \over \Gamma_1}
\, {v \over v_1} \, {f \over R} \, f_R \right)
^{(\gamma_1 - 1)/2} \bigg |_{z = z_c} \,. \eqno(5.18)
$$
At this juncture, given a streamtube function $f(R,z)$ and values
for the starting radius $R$, the starting height $z_1$, and the
starting velocity $v_1$ at the base of the wind, equations~(5.15),
(5.16), (5.17), and (5.18) can be solved simultaneously for $z_c$,
$v_c$, $s_c$, and $s_1$. Hence $s_1$ is determined by the critical
conditions once $v_1$ and $z_1$ are specified.

\medskip
\centerline{5.3. \it Corona Structure Conditions}
\medskip

In addition to the critical conditions developed in \S~5.2, we can
obtain another constraint based on the requirement that the wind joins
smoothly onto the corona at $z = z_1$, which expresses the fact that there
is no real boundary between these regions. In particular, we need to ensure
that the classical sound speed $s_1$ calculated using equation~(5.18)
is equal to the same quantity computed using the virial model that
describes the coupled disk/corona system, from which we can derive
$$
s_1^2 = {\gamma_1 \, P_1 \over N_1 \, m_p}
= {v_1 \over v_0} \, {\gamma_1 \, (\gamma_1 - 1) \, c^2\over R_* - 2} \,,
\eqno(5.19)
$$
where we have used equations~(4.1), (4.9), and (4.12) to achieve the
final result. Using equations~(4.8), (4.11), and (4.18), we can also
obtain an equation for the starting height in the corona ($z_1$) based
on the geometrical constraint $z_1 = h + H$, which yields
$$
{z_1 \over R_g} = \sqrt{R_*} \, \left[\frac{2 \, (R_*-2)^2}{R_*+2}
\sqrt{1 - {v_0 \over 2 \, v_1}}
+ \sqrt{{\gamma_0 (\gamma_0 - 1)  \, (R_* - 2) \over 2}}\right]
\,. \eqno(5.20)
$$
Treating $\xi_0 \equiv v_0 / c$ as a free parameter, equations~(5.15),
(5.16), (5.17), (5.18), (5.19), and (5.20) constitute an implicit set
of simultaneous equations for the variables $(v_1 , z_1 , s_1 , v_c ,
z_c , s_c)$ describing the critical structure of the wind originating
at cylindrical radius $R$ in the corona. Once the critical structure
is established, the vertical velocity $v$ can be calculated as a function
of the height $z$ by solving equation~(5.13) with the Bernoulli constant
set equal to $B(v_1,z_1)$. The disk model presupposes a value for $\alpha$,
and in the applications considered here we will set $\alpha = 1$ in order
to simulate the effects of rapid infall close to the black hole.

\medskip
\centerline{5.4. \it Asymptotic Lorentz Factor}
\medskip

As the energetic proton distribution emerges from the vicinity of
the central object, it cools adiabatically and expends its internal
energy by accelerating the fluid. At very large distances, the
internal energy of the protons becomes negligible due to the expansion
and therefore the sound speed $s$ vanishes. Since the gravitational
potential of the central mass also vanishes at large distances, it
follows from equation~(5.13) that once we have a critical solution,
the asymptotic Lorentz factor $\Gamma_\infty(R)$ for particles entering
the wind at cylindrical radius $R$ is equal to the Bernoulli constant,
$$
\Gamma_\infty(R) = B(v_1,z_1)
= \Gamma_1 \, \left(1 + {s_1^2 \over c^2}
{1 \over \gamma_1 - 1} \right) - {R_g \over \sqrt{R^2 + z_1^2}
- 2 \, R_g} \,. \eqno(5.21)
$$
Far from the central mass, the outflow can be regarded as a cold proton
beam moving with bulk Lorentz factor $\Gamma_\infty(R)$. Hence the
asymptotic Lorentz factor is a simple indicator of the energy
per unit mass possessed by jet particles originating at radius $R$.

\bigskip
\centerline{\bf 6. MODEL SELF-CONSISTENCY CONDITIONS}
\bigskip

We have seen that in order to escape from the vicinity of the black
hole, the wind must satisfy critical conditions that depend on
the streamtube shape, and ultimately on the nature of the collimation
mechanism. When these conditions are met, particles reach an infinite
distance with an asymptotic Lorentz factor $\Gamma_\infty$ that can
greatly exceed unity, in which case the ``wind'' has essentially
transformed into a relativistic jet. In addition to the critical
conditions, the model must also satisfy a set of self-consistency
conditions associated with (i) radial advection in the corona; (ii)
collisional losses in the corona; (iii) photon-photon pair production
in the corona; and (iv) limits on the total kinetic power carried by
the jet.

\medskip
\centerline{6.1. \it Acceleration vs. Advection}
\medskip

The protons in the corona are accelerated via collisions with kinks in
the tangled magnetic field. Since the footpoints of the tangled field
lines are embedded in the disk, they are dragged into the black hole
with the radial velocity $v_R$ given by equation~(4.7). This radial
motion is in addition to the shearing motion associated with the
variation of the azimuthal velocity $v_\phi$ given by equation~(4.2).
We must presume that the radial motion of the field lines is transmitted
to the particles in the corona along with the shear motion. For
acceleration to take place in the corona, we therefore argue that
the infall timescale for radial advection must exceed the acceleration
timescale for the second-order Fermi (shear acceleration) process. The
characteristic timescale for the shear acceleration process can be
computed by applying equation~(2.5) in the corona, yielding
$$
t_{\rm shear} \equiv {\epsilon \over \dot\epsilon_{\rm shear}}
= {1 \over 4 \, D_1} \,, \eqno(6.1)
$$
or, utilizing equations~(4.11), (4.17), and (4.18),
$$
t_{\rm shear} = {2 \, R_g \, R_*^{1/2} \, (R_* - 2)^2
\over (R_* + 2) \, v_1} \left(1 - {v_0 \over 2 \, v_1}\right)^{-1/2}
\,. \eqno(6.2)
$$
The infall timescale is given by
$$
t_{\rm infall} \equiv {R \over v_{R}}
= {2 \, R_g \, R_*^{3/2} \over \alpha \, \gamma_0 \,
(\gamma_0 - 1) \, c} \,, \eqno(6.3)
$$
where we haved used equation~(4.7) for the radial velocity $v_R$.
Combining the preceding expressions, the criterion for acceleration
to occur before the protons are swept into the
black hole becomes
$$
{t_{\rm shear} \over t_{\rm infall}}
= {Q(R_*) \over v_1} \left(1 - {v_0 \over 2 \, v_1}\right)^{-1/2}
< \ \ 1 \,, \eqno(6.4)
$$
where
$$
Q(R_*) \equiv {\alpha \, \gamma_0 \, (\gamma_0 - 1) \, c \, (R_* - 2)^2
\over R_* \, (R_* + 2)} \,. \eqno(6.5)
$$
We treat $v_0 \equiv \xi_0 c$ as a free parameter in our model, and
therefore equation~(6.4) yields a constraint on the diffusion velocity
in the corona, $v_1$, which can be expressed as
$$
v_1 > v_{\rm min} \equiv {v_0 \over 4} + \sqrt{{v_0^2 \over 16}
+ Q^2(R_*)} \,. \eqno(6.6)
$$
It is straightforward to demonstrate that $dQ/dR_* > 0$, and
therefore $v_{\rm min}$ increases with increasing $R_*$. Consequently
the outermost radius of the jet, $\Rout$,
will correspond to the largest radius at which $v_1$ exceeds
$v_{\rm min}$. We must keep this self-consistency condition
in mind when evaluating the computational results presented in \S~7.

\medskip
\centerline{6.2. \it Acceleration vs. Collisional Losses}
\medskip

In the two-temperature disks considered here, the protons are
virially hot, and they therefore collide much more frequently
with kinks in the tangled magnetic field than they do with each other
as shown in Paper 1. According to equation~(3.1), strong proton-proton
interactions are much more important than Coulomb collisions for the
energetic protons in the disk and the corona. Although rare, the
proton-proton collisions are usually catastrophic and typically
result in the near-stopping of one of the particles. Conversely,
the second-order Fermi acceleration of a proton occurs as a result
of multiple collisions with magnetic scattering centers that
inherit their shear motion from the underlying disk. In order for
net acceleration to occur in the corona, the shear acceleration timescale
expressed by equation~(6.2) must not exceed the loss timescale due
to proton-proton collisions calculated using equation~(3.2),
$$
t_{\rm loss} \equiv - {\epsilon \over \dot\epsilon_{\rm loss}}
= {1 \over \sigma_{\rm pp} \, c \, N_1} \,. \eqno(6.7)
$$
The proton number density in the corona can be evaluated
by combining equations~(4.12) and (4.21), which yields
$$
N_1 = {\dot M \, v_0 \over 4 \, \pi \, R \, H \, m_p \, v_R
\, v_1} \,. \eqno(6.8)
$$
Using equations~(6.2), (6.7), and (6.8), the criterion for net
acceleration to occur in the corona becomes
$$
{t_{\rm shear} \over t_{\rm loss}}
= {c \, \sigma_{\rm pp} \, \dot M \, v_0 \, (R_* - 2)^2 \over
2 \pi \, m_p \, H \, v_R \, v_1^2 \, (R_* + 2) \, R_*^{1/2}}
\left(1 - {v_0 \over 2 \, v_1}\right)^{-1/2}
< \ \ 1 \,. \eqno(6.9)
$$
Substituting for $v_R$ and $H$ using equations~(4.7) and (4.8),
respectively, yields an expression for the maximum accretion rate
as a function of radius, which can be written as
$$
{\dot M_{\rm max}(R_*) \over \dot M_{\rm E}} \equiv
{\alpha \, 5^{3/2} \over 54} \,
{\sig \over \sigma_{\rm pp}} \,
{v_1^2 \over v_0 c} \, {R_* + 2 \over R_*}
\left(1 - {v_0 \over 2 \, v_1}\right)^{1/2}
\left(1 - {2 \over R_*}\right)^{-3/2} \,, \eqno(6.10)
$$
where $\dot M_{\rm E}\equiv L_{\rm E}/c^2$ is the Eddington accretion
rate corresponding to the Eddington luminosity $L_{\rm E}\equiv
4 \pi G M m_p c / \sig$. Note that the value of
$\dot M_{\rm max}$ at a particular radius $R_*$ also depends
on the value of $v_1$ obtained at that radius.
In specific applications, we generally find that $\dot M_{\rm max}
\gg \dot M_{\rm E}$, and therefore collisional losses do not provide
a very strong constraint on the model.

\medskip
\centerline{6.3. \it Pair Production in the Corona}
\medskip

The detailed results presented in \S~7 suggest that energy losses
due to strong interactions are unimportant for the protons in the
corona unless the accretion rate is highly super-Eddington. We
established in \S~3 that losses due to strong interactions
dominate over proton-proton Coulomb losses in the corona, and
these in turn dominate over proton-electron Coulomb losses
(Schmidt 1966), implying that proton-electron Coulomb interactions
are negligible in the corona. However, this conclusion must
be modified if electron-positron pair production takes place at a
significant rate in the corona, because the presence of pairs
can substantially reduce the timescale for the protons to cool
via Coulomb coupling.

We can estimate the density of electron-positron pairs in the corona
by considering the probability that a $\gamma$-ray of energy $E_\gamma$
will be converted into an electron-positron pair while traversing the
corona due to a collision with another (target)
$\gamma$-ray. For simplicity, we assume that the target $\gamma$-ray
also has energy $E_\gamma$. The resulting self-interaction optical
depth can be calculated using equation~(6.4) from Becker \& Kafatos
(1995), which yields
$$
\tau_{\gamma\gamma}(E_\gamma) = 8 \, {m_p \over m_e} \,
{L_\gamma \over L_{\rm E}} \, \left(h \over R_g\right)^{-1}
\, \left(E_\gamma \over m_e \, c^2\right)^{-5} \, \int_0^{\beta_{\rm max}}
{\sigma_{\gamma\gamma}(\beta) \over \sig} \, {2 \, \beta \,
d\beta \over (1 - \beta^2)^3} \,, \eqno(6.11)
$$
where $h$ is the thickness of the corona, $L_\gamma$ is the
$\gamma$-ray luminosity,
$$
\beta_{\rm max} = \left[1 - \left(E_\gamma \over m_e \, c^2\right)
^{-2}\right]^{1/2} \,, \eqno(6.12)
$$
and
$$
\sigma_{\gamma\gamma}(\beta) = {3 \over 16} \, \sig \, (1 - \beta^2)
\, \left[(3 - \beta^4) \, \ln\left(1 + \beta \over 1 - \beta\right)
- 2 \, \beta \, (2 - \beta^2)\right] \,. \eqno(6.13)
$$
For typical $\gamma$-ray blazars, most of the observed luminosity
is carried by photons with energy $E_\gamma \sim 1\,$GeV, in which
case we obtain
$$
\tau_{\gamma\gamma} \approx 10^{-5} \, {L_\gamma \over L_{\rm E}}
\left(h \over R_g\right)^{-1} \,. \eqno(6.14)
$$
The detailed results presented in \S~7 indicate that $h/R_g \sim 10$,
and we therefore conclude that photon-photon pair production in the corona
is unimportant for any reasonable value of $L_\gamma$. We can extend this
argument to conclude that pair production via particle collisions is also
insignificant. It follows that no pair-related enhancement in the
proton-electron Coulomb loss rate occurs, and consequently our assumption
of negligible Coulomb losses in the corona is justified.

\medskip
\centerline{6.4. \it Asymptotic Kinetic Power}
\medskip

We can derive another type of model constraint by computing the
asymptotic kinetic power of one of the jets in comparison to
the theoretical limit $(1/2) \, \Mout \, c^2$ for cold matter
accreting from rest at infinity. At a large distance
from the central mass, the internal energy of the proton
distribution becomes negligible compared with the kinetic energy
of the bulk motion, and the asymptotic kinetic power of one of
the jets can therefore be calculated using
$$
L_{\rm jet}(R) = - c^2 \, \int_R^{\Rout} \Gamma_\infty(R') \,
\frac{d \, \dot m(R') }{dR'} \, dR' \, \sim \, {\rm ergs \ s^{-1}}
\,, \eqno(6.15)
$$
which sums up the contributions due to energetic protons feeding
the base of the outflow between radii $R$ and $\Rout$. The negative
sign arises because $d \dot m / dR < 0$. We can use
equations~(4.19) and (4.22) to rewrite equation~(6.15) as
$$
L_{\rm jet}(R) = {c^2 \over 2 \, \sqrt{R_g}} \,
\left(9 \over 5\right)^{3/2} \, {\xi_0 \over \alpha} \,
\int_R^{\Rout} {\Gamma_\infty(R') \, \dot M(R') \over
\sqrt{R' - 2 \, R_g}} \, dR' \,, \eqno(6.16)
$$
where $\dot M$ is evaluated using equation~(4.23).

Once a series of critical solutions has been obtained for the
range of radii contained within the bounds of integration
and the corresponding values of $\Gamma_\infty(R)$ have been
computed using equation~(5.21), equation~(6.16) can be used to
calculate the kinetic luminosity of the jet. Note that
$L_{\rm jet}$ diverges as the horizon is approached
($R \to 2 R_g$) because of the behavior of the denominator
in equation~(6.16). We can therefore establish a rough estimate
for the innermost radius of the jet, $\Rin$, by setting
$$
L_{\rm jet}(\Rin) = {1 \over 2} \, \Mout c^2 \,. \eqno(6.17)
$$
In the vicinity of $\Rin$, we expect the validity of our model
to break down because the disk has lost virtually all of its internal
energy to the wind/jet, and therefore our assumption of virial disk
structure obviously becomes inconsistent. Nonetheless, the values
obtained for $\Rin$ in specific models will provide useful insight
into the radial extent of the jet. With $\Rin$ determined using
equation~(6.17), it is interesting to compute the mean asymptotic
Lorentz factor of the jet by averaging $\Gamma_\infty(R)$ between
$\Rin$ and $\Rout$, weighted by the
differential mass loss rate. The mean value obtained is
$$
\langle \Gamma_\infty \rangle \equiv - {1 \over \dot m(\Rin)}
\int_{\Rin}^{\Rout} \Gamma_\infty(R') \,
\frac{d \, \dot m(R') }{dR'} \, dR'
= {1 \over 2} \, {\Mout \over \dot m(\Rin)}\,, \eqno(6.18)
$$
where the final result follows from equations~(6.15) and (6.17).
We can substitute for $\dot m(\Rin)$ using equation~(4.24) to
obtain the equivalent result
$$
\langle \Gamma_\infty \rangle = {\Mout \over \Mout - \Min}\,, \eqno(6.19)
$$
where $\Min \equiv \dot M(\Rin)$ is the rate at which matter
actually crosses the event horizon and enters the hole.

\vfil
\eject

\bigskip
\centerline{\bf 7. RESULTS}
\bigskip

The rate at which the proton distribution cools and deposits its
energy into the bulk motion of the plasma depends on the shape of
the streamlines. As explained
in \S~5, we arrive at a one-dimensional problem by assuming azimuthal
symmetry in the outflow and introducing the streamtube function
$r \equiv f(R, \, z)$, which is the locus of streamlines originating
at a given radius $R$ in the corona. All of the critical point
constraints described in \S~5 depend on
the streamtube function, which has been left unspecified up to
this point. The actual shape of the streamlines may be very complicated,
and there have been many attempts to understand the collimation
of the flow on various scales (e.g., Blandford \& Payne 1982; Contopulos
\& Lovelace 1994; Lynden-Bell 1996). In view of the prevalent uncertainty,
we adopt the conical outflow model utilized by Mannheim (1993) and by
Blandford \& Konigl (1979),
$$
r = f(R, \, z) = R + (z - z_{1}) \, {\rm tan} \, \theta \,, \eqno(7.1)
$$
where $\theta$ is the half-angle of the flow.

We next discuss representative results obtained using four models
constructed using two values each for the half-angle $\theta$ and
the tangling parameter $\xi_0$. Once the global parameters $\theta$,
$\xi_0 \equiv v_0/c$, and $\alpha$ are specified, we determine the
critical structure of the coupled disk/corona/wind system by solving
simultaneously equations~(5.15), (5.16), (5.17), (5.18), (5.19), and
(5.20) for the quantities $(v_1 , z_1 , s_1 , v_c , z_c , s_c)$ as
functions of the (cylindrical) starting radius in the corona $R$.
Once the critical structure has been established, we can solve for
the variation of the vertical velocity $v$ as a function of the height
$z$ by solving equation~(5.13) as discussed in \S~5. We set $\alpha = 1$
in all of the models in order to simulate the effects of rapid infall
between the radius of marginal stability and the event horizon. In Models 1 and 2
we set $\xi_0 = 0.05$, and in Models 3 and 4 we set $\xi_0 = 0.1$.
In Models 1 and 3 we set $\theta = 0.5^\circ$, and in Models 2 and 4
we set $\theta = 1.0^\circ$. Our graphical results will be presented
using a uniform set of line styles for Model 1 ({\it solid line}),
Model 2 ({\it dotted line}), Model 3 ({\it dashed line}), and Model 4
({\it dot-dashed line}). In Table~1 we summarize the
results obtained in each case for the inner
jet radius $\Rin$, the outer jet radius $\Rout$, the mean asymptotic
Lorentz factor $\langle\Gamma_\infty\rangle$, and the accreted mass fraction
$\Min/\Mout$.
The corresponding results obtained for the critical parameters
$(v_1 , z_1 , s_1 , v_c , z_c , s_c)$ at the jet turn-on radius
$R = \Rout$ are presented in Table~2 for each of the models.

In Figure~3 we plot the velocity ratio $v_1 / v_0 = \xi_1 / \xi_0
= N_0 / N_1$ (see eqs.~[4.11] and [4.13]) as a function of the
dimensionless cylindrical starting radius $R_* \equiv R c^2 / G M$
for each of the four models. We also plot curves corresponding to
$v_{\rm min} / v_0$ computed using equation~(6.6). As discussed in \S~6,
when $v_1 < v_{\rm min}$, protons are advected into the black
hole before they can be accelerated via the Fermi process, and
therefore the condition $v_1 = v_{\rm min}$ defines the outermost
radius of the jet, $\Rout$ (see Table~1). At radii smaller than
$\Rout$, acceleration dominates over advection because of the steep
nature of the pseudo-Newtonian potential. The plots indicate that
$N_1 < N_0$, which is reasonable since the corona is populated by
protons that diffuse out of the disk. Note that the velocity ratio
essentially indicates the ratio of the mean proton energies in the
corona and the disk, since
$$
{\langle \epsilon_1 \rangle \over \langle \epsilon_0 \rangle}
= {1 \over 1 - y} = 2 \, {v_1 \over v_0} \,, \eqno(7.2)
$$
which we have obtained by combining equations~(3.9) and (4.16).
Figure~3 therefore implies that $\langle \epsilon_1 \rangle /
\langle \epsilon_0 \rangle \sim 8 - 20$ for the computed models,
or, equivalently, $y \sim 0.88 - 0.95$.

In Figure~4 we depict the results obtained for the coherence
length in the corona divided by the coherence length in the
disk, $\tilde\lambda_1 / \tilde\lambda_0$, as a function of $R_*$.
We see that this ratio always exceeds unity in all of the plots,
indicating that the field lines are less tangled in the corona
than they are in the disk. This is consistent with the simulations
of Romanova et al. (1998), who found that magnetic loops in the corona
tend to open up due to differential rotation in the
disk. In Figure~5 we plot the solutions obtained for the height of the
wind-corona interface, $z_1$, and for the height of the critical surface,
$z_c$, as functions of the starting radius $R_*$ for each model. The
height of the
critical surface increases with increasing radius due to the decrease
in the depth of the gravitational potential at the base of the flow.
Note that for a fixed value of $\xi_0$, the critical surface moves away
from the disk as $\theta$ decreases. In the limit of cylindrical
flow ($\theta \to 0$), the critical surface is pushed out to infinity,
and the flow loses its critical behavior
entirely. The critical height $z_c$ generally lies above
the starting height $z_1$, in which case the flow is subsonic at
its base. However, the behavior of Model 2 is reversed
in the sense that the flow is already supersonic
at the base (i.e., $z_c < z_1$). In this case, the critical point
is virtual in nature, and does not actually exist in the flow.

In Figure~6 we plot the asymptotic Lorentz factor $\Gamma_\infty$
(eq.~[5.21]) as a function of $R_*$ for each of the models. In all
four cases $\Gamma_\infty$ greatly exceeds unity in the inner region,
indicating the formation of a relativistic jet. The asymptotic Lorentz
factor tends to decrease with increasing radius due to the diminishing
strength of the Keplerian shear. For a fixed value of $\theta$,
$\Gamma_\infty$ tends to increase with decreasing $\xi_0$, reflecting
an increase in the sound speed at the base of the flow (cf. Fig.~3 and
eq.~[5.19]). In general, we find that $2 \lapprox \Gamma_\infty\lapprox 10$,
in reasonable agreement with observations of the bulk Lorentz factors of
blazar jets (e.g., Vermeulen \& Cohen 1994; von Montigny et al. 1995).

In Figure~7 we use equation~(4.23) to plot the radial variation of
the disk accretion rate $\dot M$ as a function of $R_*$
for each of the models. The accretion rate decreases with decreasing
radius in response to the loss of mass into the jet and the counterjet.
The curves are normalized with respect to the accretion rate $\Mout$
at the jet turn-on radius $\Rout$ (see Fig.~3 and Table~1), and the
accreted mass fraction corresponds to $\dot M(\Rin)/\Mout$.
In Figure~8 we plot the upper limit for the accretion rate,
$\dot M_{\rm max}(R_*)/\dot M_{\rm E}$, evaluated using equation~(6.10).
When the actual accretion rate $\dot M$ exceeds $\dot M_{\rm max}$,
losses due to proton-proton collisions in the corona dominate
over Fermi energization, and the protons are decelerated rather than
accelerated. It is apparent from Figure~8 that jets can exist provided
$\Mout / \dot M_{\rm E} \lapprox 10-20$, so that the accretion rate
supplied to the outer edge of the disk is not very strongly constrained,
and the accretion can be highly super-Eddington. However, for accretion rates
in excess of $\dot M_{\rm E}$, the underlying disk
will cool so much that it loses its two-temperature character,
resulting in the turning off of the jet (Rees et al. 1982).

In Figure~9 we use equation~(6.16) to plot the asymptotic kinetic
luminosity of the jet, $L_{\rm jet}(R)$, divided by the maximum
accretion luminosity $(1/2) \, \Mout \, c^2$. The inner radius of
the jet, $\Rin$, is defined as the radius at which this ratio reaches
unity (see eq.~[6.17] and Table~1). The estimates obtained for $\Rin$
in this manner are obviously quite rough, since the actual kinetic
luminosity is not likely to approach $(1/2) \, \Mout \, c^2$.

In order to investigate the vertical structure of the flow for a
particular value of the starting radius $R$, in Figure~10 we plot the
Lorentz factor $\Gamma$ computed using equation~(5.7) as a function of
the height $z$ measured from the midplane of the disk. In each case we
set $R = \Rout$ so that the results describe the outer edge of the jet.
It is apparent that the flow accelerates strongly out to a
distance of $\sim 10^6 \, R_g$ from the central mass, which corresponds
to $\sim 100 \,$pc for a $10^9 \, \Msun$ black hole. On the other hand,
at a distance of $\sim 1 \, {\rm pc} \sim 10^4 \, R_g$, the jet may very
well collide with a broad-line cloud and become disrupted, resulting in
the production of high-energy $\gamma$-ray emission. The asymptotic
Lorentz factor $\Gamma_\infty$ has not yet been achieved at this distance,
but the flow is nonetheless highly relativistic ($\Gamma \gapprox 2$),
and therefore most of the energy flux (both kinetic and internal)
should be converted into observable $\gamma$-rays as we discuss in
\S~8.

We further explore the transonic nature of the flow in Figure~11
by plotting the Mach number for the motion in the $(r,z)$ plane,
$$
{\cal M}_{rz} \equiv {v \, (1 + f_z^2)^{1/2} \over a} \eqno(7.3)
$$
as a function of $z$,
where $a$ is the relativistic sound speed (see eq.~[5.12]). As in
Figure~10, we set $R = \Rout$. The sonic transition occurs at $z = z_c$,
where ${\cal M}_{rz} = 1$. Essentially all of the acceleration occurs
beyond the sonic point, which acts as a ``throttle'' in the flow.
Finally, in Figure~12 we plot curves representing
the variation of the number density in the wind, $N$, as a function
of $z$ for $R = \Rout$. The density declines sharply above the base
of the flow at $z = z_1$, suggesting that shear (viscous) stresses
will not have a strong effect on the structure of the wind/jet.

\bigskip
\centerline{\bf 8. DISCUSSION AND CONCLUSION}
\bigskip

We have investigated the properties of a coupled disk-corona
model in which the corona joins onto a relativistic proton-electron
wind, which transforms into a jet far from the central object. The
expansion of the wind is powered by the pressure of the protons,
which are accelerated in the corona via a second-order Fermi
mechanism driven by the shear flow in the underlying accretion disk.
As discussed in \S~3, this mechanism preferentially accelerates
high-energy protons and consequently it adds a power-law tail to
the thermal proton distribution diffusing into the corona from
the disk. The acceleration of the relativistic protons occurs
at the expense of the azimuthal kinetic energy of the disk, and
in this sense the entire process acts as a source of viscosity
in the disk. This viscosity is essentially magnetic in nature,
and the dissipated energy is carried from the disk into the corona
via the Poynting flux of the shearing magnetic field.

The shear-driven Fermi acceleration
process transfers energy to all of the protons, whether they are
located in the disk or the corona. However, the consequences of
the energy transfer differ depending on the importance of
proton-proton collisions, which tend to thermalize the energy.
According to our results, the density in the corona is low enough
to facilitate the development of a power-law tail, whereas in the
disk collisional thermalization is
probably efficient enough to maintain a Maxwellian proton distribution.
The transfer of energy to the thermal disk protons
via the Fermi acceleration process is equivalent to conventional
viscous heating, and the energy therefore changes form (from
kinetic to internal) but still remains in the disk.

We have chosen to focus on hot, two-temperature accretion disks around
black holes in this paper because such disks are an abundant source of
high-energy seed protons. Motivated by the results of MHD simulations
performed by various groups, we have assumed that the corona is threaded
by a tangled magnetic field which is frozen into the underlying disk.
The disk structure is treated in an approximate manner by assuming
a virial value for the disk energy density. The connection between the
structures of the disk and the corona is established by assuming that
vertically propagating sound waves enforce pressure equilibrium between
the two regions. We are thus able to establish a clear link
between the accretion disk and the relativistic outflow which is
presumably responsible for many of the observed properties of blazars.
In particular, proton-dominated jets have been invoked frequently
as a likely explanation for the observed GeV and TeV emission from
blazars, and this work therefore represents an important step in building
a complete picture that couples the central black hole, the
surrounding accretion disk, and the relativistic outflow.

The general conclusion that can be drawn from our results
is that relativistic proton outflows with asymptotic Lorentz
factors $\Gamma_\infty \lapprox 10$ can exist so long as the
gas is supplied at a rate $\Mout / \dot M_{\rm E} \lapprox 10 - 20$.
This large upper limit implies that relativistic outflows can
exist for accretion rates
spanning the range from sub-Eddington to highly super-Eddington,
although the existence of the underlying hot accretion disk may become
questionable for substantially super-Eddington accretion rates due
to efficient bremsstrahlung cooling.
The values we obtain for the asymptotic Lorentz factor are consistent
with those implied by observations of blazars (Vermeulen \& Cohen 1994;
von Montigny et al. 1995), although it is not clear that the flows
considered here actually achieve the asymptotic Lorentz factor, since
this may require the jet to remain intact up to $10 \,$kpc from the
central object.

\medskip
\centerline{8.1. \it Production of Radiation in Jet-Cloud Collisions}
\medskip

Given the uncertainty in the geometry of the jet, the actual
division of energy between internal and kinetic forms that
characterizes the jet at the time it collides
with target protons in a broad-line emission cloud is unclear.
It is therefore difficult to predict the detailed spectrum
of the emergent $\gamma$-radiation for this process. However,
the asymptotic Lorentz factor of the jet provides some overall
indications. Let us assume for example that the jet has
$\Gamma_\infty \sim 10$, and that it is intercepted by a
broad-line emission cloud after it has achieved
its asymptotic Lorentz factor. The average energy of the emergent
radiation in a proton-proton reaction is $\sim 0.1 \times 0.5$ times the
energy of the incident high-energy proton involved in the reaction
(Katz 1991). The factor of 0.1 expresses the overall efficiency of the
cascade process, and the factor of 0.5 appears because two photons
will emerge from a reaction initiated by a single proton. Since the
total energy carried by a cold proton moving with $\Gamma_\infty \sim 10$
is $\sim 10$~GeV, we conclude that radiation from the process described
above will be centered around 0.5~GeV, and that the total luminosity
of the radiation will be $\sim 5 \,$\% of the asymptotic kinetic power
of the jet. If the asymptotic Lorentz factor has not been achieved
by the time the jet collides with the cloud, then a significant
fraction of the jet luminosity is still carried in the form of
advected internal energy. However, the results are not likely to
be very different in this case since the total power of the jet
is nearly equal to the asymptotic power at large distances from
the black hole.

We can estimate the kinetic jet luminosity required to power
an observed flux of $\gamma$-rays by employing a simple model
for the emission geometry describing the angular distribution
of the radiation produced in the cascade. When a conical jet
of the type considered in \S~7 collides with a broad-line cloud,
we expect the cascade radiation to be emitted in a conical distribution
with an opening angle $\Phi$ given roughly by
$$
\Phi \approx \theta + \langle \Gamma_\infty \rangle^{-1} \,. \eqno(8.1)
$$
The flux of the observed $\gamma$-radiation is related to
the $\gamma$-ray luminosity by
$$
F_{\rm rad} = {L_{\rm rad} \over \Omega \, {\cal L}^2} \,, \eqno(8.2)
$$
where ${\cal L}$ is the distance from the observer to the source and
$\Omega$ is the solid angle covered by the emitted radiation.
For small values of $\Phi$, we have $\Omega \approx \pi \, \Phi^2$,
and equations~(8.1) and (8.2) can therefore be combined to obtain
$$
L_{\rm rad} \approx \pi \, {\cal L}^2 \, F_{\rm rad} \,
\left(\theta + \langle \Gamma_\infty \rangle^{-1}\right)^2 \,.
\eqno(8.3)
$$

In the case of the June 1993 flare of 3C~279 observed by {\sl EGRET} on
the {\sl Compton Gamma-Ray Observatory}, we can use ${\cal L} = 5.65
\times 10^{27} \,$cm and $F_{\rm rad} = 2.49 \times 10^{-9}\,{\rm ergs}
\ {\rm sec}^{-1}\,{\rm cm}^{-2}$ (Becker and Kafatos 1995; Hartman 1992)
along with the results for $\langle \Gamma_\infty \rangle$ in Table~1 to obtain
$L_{\rm rad} / 10^{46}\,{\rm ergs\ s}^{-1} \approx 0.73,\, 0.88,\,
1.40,\,$ and 1.70 for Models 1, 2, 3, and 4, respectively. The
corresponding values for the kinetic luminosity of the jet are $L_{\rm
jet} / 10^{47}\,{\rm ergs\ s}^{-1} \approx 1.47,\, 1.76,\, 2.80,\,$ and
3.40, respectively, computed by setting $L_{\rm rad} = 0.05 \, L_{\rm
jet}$. These results for $L_{\rm jet}$ are comparable to the Eddington
luminosity for a $10^9 \, \Msun$ black hole. Beall \& Bednarek (1999)
have recently considered a model in which a proton beam produces
$\gamma$-rays upon collision with a broad-line cloud via pion
production. The protons in their jet model have a power-law spectrum
with an index that is set arbitrarily. Within the context of our model,
the power-law index can be calculated using equation~(3.14), and
therefore a unified description of the jet can be constructed that
connects the acceleration in the corona with the ultimate production
of the $\gamma$-rays.

\medskip
\centerline{8.2. \it Shear Acceleration vs. Stochastic Acceleration}
\medskip

We have demonstrated that collisions with cold magnetic scattering
centers embedded in a Keplerian shear flow provide an acceleration
mechanism for the protons that is formally equivalent to a second-order
Fermi process. However, in reality the scattering centers will be
propagating MHD waves rather than cold kinks. The wave aspect imparts a
{\it stochastic} component of motion to the scattering centers, which
will tend to augment the particle acceleration associated with their
shear motion. It is therefore important to determine which mechanism
dominates in a typical situation. The relative strengths of the two
processes can be estimated by comparing their respective acceleration
rates.

The stochastic motion of the waves will energize the protons via
second-order Fermi acceleration, and in this sense it is similar
to the shear acceleration mechanism. The stochastic process can
therefore also be modeled using equation~(2.2), and the mean
acceleration rate is consequently given by equation~(2.5),
$$
\langle \dot\epsilon_{\rm stoch} \rangle = \frac{1}{\epsilon^2}
\frac{d}{d \epsilon} \left( \epsilon^2 {\cal D}_{\rm stoch} \right)
\,, \eqno(8.4)
$$
where ${\cal D}_{\rm stoch}$ is the energy diffusion coefficient
for the stochastic acceleration process. The spatial diffusion coefficient
in the corona, $\kappa_1$, is related to ${\cal D}_{\rm stoch}$ via
(e.g., Becker, Kafatos, \& Maisack 1994)
$$
\kappa_1 \, {\cal D}_{\rm stoch} = {v_{\rm A}^2 \, \epsilon^2
\over 9} \,, \eqno(8.5)
$$
where $v_{\rm A} = B_1 / \sqrt{4 \pi \rho_1}$ is the Alfv\'en velocity
in the corona associated with magnetic field strength $B_1$ and
density $\rho_1$. Combining equations~(8.4) and (8.5) and
substituting for $\kappa_1$ using equation~(3.6) yields
$$
\langle \dot\epsilon_{\rm stoch} \rangle = {B_1^2 \, \epsilon
\over 3 \, \pi \, c \, \rho_1 \, \tilde\lambda_1} \,, \eqno(8.6)
$$
where we have assumed that $\kappa_1$ is independent of $\epsilon$.
Defining $B_0$ as the magnetic field strength in the disk and using
equation~(4.1), we obtain
$$
\langle \dot\epsilon_{\rm stoch} \rangle = {4 \over 3} \,
\left(B_1 \over B_0\right)^2 \, {\rho_0 \over \rho_1} \,
{c \, \epsilon \over \beta_0 \, \tilde\lambda_1} \,  \,
{1 \over R_* - 2} \,, \eqno(8.7)
$$
where $\beta_0 \equiv 8 \, \pi \, U_0 / B_0^2$ is the plasma
$\beta$-parameter in the disk.

We can now evaluate the ratio
of the shear and stochastic acceleration rates by combining
equations~(2.5), (4.17), (4.18), and (8.7), which yields
$$
{\langle \dot\epsilon_{\rm shear} \rangle \over
\langle \dot\epsilon_{\rm stoch} \rangle}
= {3 \over 4} \, \left(v_0 \over c\right)^2 \,
\left(B_0 \over B_1\right)^2 \, \beta_0 \,
(R_* - 2) \, \left({v_1 \over v_0} - {1 \over 2}\right) \,.
\eqno(8.8)
$$
Simulations of MHD dynamos driven by turbulence in accretion disks
suggest that the strength of the magnetic field in the saturated state
corresponds to $\beta_0 \sim 100$ (e.g., Hawley, Gammie, \& Balbus 1995, Brandenburg
et al. 1995). We adopt this value in Figure~13, where we use equation~(8.8)
to plot the ratio $\langle \dot\epsilon_{\rm shear} \rangle / \langle
\dot\epsilon_{\rm stoch} \rangle$ as a function of the dimensionless
starting radius in the corona $R_*$ for each of the four models considered
in \S~7. We expect that $B_0 \gapprox B_1$ since the field is generated
by the shear in the disk. However, in order to obtain a conservative
estimate, we set $B_0 = B_1$ when performing the calculations plotted
in Figure~13. Note that the shear acceleration rate greatly exceeds that
due to the stochastic wave motion at all radii, and consequently we have
not committed a serious error by neglecting the latter process in our
analysis of the particle acceleration occurring in the corona.

We have shown that the relativistic protons in the corona experience
negligible losses due to collisions with other protons. However, the
coronal protons also cool via the excitation of MHD waves which are
amplified as a consequence of the super-Alfv\'enic motion of the protons
along the field lines. If this cooling mechanism imposes losses
comparable to the rate of shear-driven Fermi acceleration, then the net
acceleration rate will be substantially lower than the value used in our
model. This possibility can be ruled out by carefully considering the
rates for the various energy transfer processes as follows. In a
saturated steady state, the rate at which the waves transfer energy to
the protons via stochastic acceleration is equal to the rate at which
the protons transfer energy to the waves via the excitation process {\it
plus} a portion of the Poynting flux from the disk (the remainder of the
Poynting flux generates the shear motion of the waves). This implies
that the rate at which the protons lose energy due to wave excitation is
less than the rate at which the protons gain energy from the waves via
stochastic acceleration. According to Figure~13, the latter process is
negligible compared with the shear-driven Fermi acceleration rate, and
therefore it follows that the cooling of the protons via the excitation
of MHD waves is also insignificant compared with the rate of shear
acceleration as we have assumed.

\medskip
\centerline{8.3. \it Conclusion}
\medskip

In this paper we have demonstrated the plausibility of the production
of relativistic jets as a direct consequence of the accretion of
material onto a supermassive black hole. The asymptotic bulk Lorentz
factors obtained in our models are commensurate with those inferred
from observations of superluminal motion in blazars (Vermeulen \& Cohen
1994). Having demonstrated the basic
framework, it is now possible to build detailed models that can predict
both the spectrum and the bulk Lorentz factor of the energetic proton
population in the jet at an arbitrary distance from the central object.
This information can then be used as input to detailed calculations
of proton-initiated reaction cascades that ultimately result in
observable high-energy emission.

Although we have focused on hot, two-temperature accretion disks as the
source of particles and energy for the jet in this paper, the
relativistic wind formalism developed in \S~5 is not specific to the
disk model used here. Advection-dominated accretion flows (Narayan \& Yi
1995) represent another interesting alternative, since these flows have
the attractive properties of being hot and tenuous, and may prove to be
natural environments for the shear acceleration mechanism studied here.
Blandford \& Begelman (1999) have outlined a generalization of the
advection-dominated model that includes outflows which carry away some
of the binding energy of the infalling plasma. Such outflows alleviate
the problem of the accreting gas having a net positive energy and
therefore being gravitationally unbound, although no mechanism for
generating the outflows is specified by the authors. It is conceivable
that the shear acceleration mechanism described in this work could power
the outflows.

The protons in the jet are expected to drag electrons along
with them via Coulomb coupling, but we do not expect the inertia of the
electrons to load down the jet to any appreciable extent because
the electrons are much less massive than the protons. However, the
cooling of the electrons via synchrotron and inverse-Compton radiation
presents a potentially more serious problem, since this may lead to the
cooling of the protons which would tend to quench the expansion. The
adiabatic losses considered in our model take place over a sound
crossing timescale, which is roughly equal to the dynamical timescale.
The dynamical timescale is in turn expected to be far smaller than
the electron-ion Coulomb coupling timescale since the density drops
very rapidly above the disk (see Fig.~12). Hence it is not likely that
electron-ion coupling will be able to cool the protons substantially
before the jet has expanded to the point where the electrons and protons
have essentially stopped interacting. This may lead to the development
of a charge separation in the jet. We also note that a more efficient
form of electron-ion coupling may occur as a result of collective
modes (Begelman \& Chiueh 1988).

Strictly speaking, the distribution function and the bulk Lorentz
factor $\Gamma$ should be determined self-consistently along with
the flow velocity, and this is the approach taken by Contopoulos \&
Kazanas (1995). In such a treatment, the adiabatic index of the wind
cannot be regarded as a constant, and must be calculated at every
spatial location. It typically varies from $4/3$ for the highly
relativistic gas at the base of the outflow up to $5/3$ for the cold,
nonrelativistic gas at large distances from the central object. Here
we have chosen to avoid the complexity of dealing with a variable
adiabatic index, and instead we have used the corona value $\gamma_1
= 4/3$ to describe the thermodynamics throughout the entire wind.
The validity of this procedure can be evaluated a posteriori by
using the comoving energy spectrum of the protons to recalculate the
adiabatic index. We plan to compute the comoving proton
distribution and the detailed $\gamma$-ray spectrum resulting from
a jet-cloud collision in a later paper.

We would like to thank Professor Menas Kafatos for numerous insightful
discussions throughout the development of this work.

\vfil
\eject

\centerline{\bf APPENDIX A}
\centerline{\bf CALCULATION OF THE MEAN ACCELERATION RATE}
\bigskip

In this section we establish that the mean shear acceleration rate
for particles with energy $\epsilon$ associated with the transport
equation
$$
\frac{\partial f}{\partial t} = - \frac{1}{\epsilon^2}
\frac{\partial}{\partial \epsilon} \left( - \epsilon^2 \, {\cal D}
\frac{\partial f}{\partial \epsilon} \right) \eqno({\rm A}1)
$$
is given by
$$
\langle \dot{\epsilon} \rangle_{\rm shear} = \frac{1}{\epsilon^2}
\frac{d}{d \epsilon} \left( \epsilon^2 {\cal D} \right)
\,, \eqno({\rm A}2)
$$
as quoted in \S~2. We begin by defining the energy density
$$
U(t) \, = \int_0^\infty \epsilon^3 \,
f(\epsilon,t) \, d \epsilon \, \, \sim \ {\rm ergs \ cm^{-3}} \,,
\eqno({\rm A}3)
$$
and the number density
$$
N(t) \, = \int_0^\infty \epsilon^2
f(\epsilon,t) \, d \epsilon \, \, \sim \ {\rm cm^{-3}} \,.
\eqno({\rm A}4)
$$
Note the arguments made following equation~(2.4) in reference to
the lower bounds of the integrals in equations~(A3) and (A4).
Defining the average energy per particle as
$$
\langle \epsilon \rangle = \frac{U}{N} \,, \eqno({\rm A}5)
$$
we find that the rate of change of $\langle \epsilon \rangle$
is given by
$$
\frac{d}{dt} \langle \epsilon \rangle =
\frac{N \, dU/dt - U \, dN/dt}{N^2} \,.
\eqno({\rm A}6)
$$
Let us focus on the evolution of $J$ particles per unit volume
all having energy $\epsilon = \epsilon_i$ at time $t = t_i$. If
$f(\epsilon,t)$ denotes their distribution, then at time $t = t_i$
we have
$$
f(\epsilon, t) \, \Big|_{t = t_i} = {J \over \epsilon_i^2} \,
\delta (\epsilon - \epsilon_i) \,, \eqno({\rm A}7)
$$
so that the particle number density computed using equation~(A4)
is $J$ as required. We can combine equation~(A4) with the kinetic
equation~(A1) to write the rate of change of the number density as
$$
{dN \over dt} = \int_0^\infty \epsilon^2 \, {\partial f \over
\partial t} \, d\epsilon
= \int_0^\infty {\partial \over \partial \epsilon}
\left(\epsilon^2 \, {\cal D} \, {\partial f \over \partial \epsilon}
\right) d \epsilon \,. \eqno({\rm A}8)
$$
At time $t=t_i$, the initial condition given by equation~(A7) can
be employed to obtain
$$
{dN \over dt} \, \bigg|_{t = t_i}
= \lim_{t \to t_i} \ \ \epsilon^2 \, {\cal D} \frac{\partial}
{\partial \epsilon}
f(\epsilon,t) \, \bigg|_0^\infty = 0 \,, \, \, \eqno({\rm A}9)
$$
so that there is no instantaneous change in the particle number
density. The analog of equation~(A8) for the energy density $U$
is
$$
{dU \over dt} = \int_0^\infty \epsilon^3 \, {\partial f \over
\partial t} \, d\epsilon
= \int_0^\infty \epsilon \, {\partial \over \partial \epsilon}
\left(\epsilon^2 \, {\cal D} \, {\partial f \over \partial \epsilon}
\right) d \epsilon \,. \eqno({\rm A}10)
$$
Integrating by parts twice yields
$$
{dU \over dt} = \epsilon^3 \, {\cal D} \, {\partial f \over
\partial \epsilon} \bigg|_0^\infty
- \epsilon^2 \, {\cal D} \, f \bigg|_0^\infty
+ \int_0^\infty {\partial \over \partial \epsilon}
\left(\epsilon^2 \, {\cal D} \right) \, f \, d\epsilon \,,
\eqno({\rm A}11)
$$
or, incorporating the initial condition given by equation~(A7),
$$
{dU \over dt} \, \bigg|_{t = t_i}
= {J \over \epsilon^2} \, {\partial \over \partial \epsilon}
\left( \epsilon^2 \, {\cal D} \right) \bigg|_{\epsilon=\epsilon_i}
\,. \eqno({\rm A}12)
$$
Combining equations (A6), (A9), and (A12), we arrive at
$$
\frac{d}{dt} \langle \epsilon \rangle \, \bigg|_{t = t_i}
= {1 \over \epsilon^2} \, {\partial \over \partial \epsilon}
\left( \epsilon^2 \, {\cal D} \right) \bigg|_{\epsilon=\epsilon_i}
\,. \eqno({\rm A}13)
$$
Although we have treated particles with energy $\epsilon = \epsilon_i$
at time $t=t_i$, the result is valid for particles with any
chosen energy at any time, so that we obtain
$$
\frac{d\langle \epsilon \rangle}{dt}
= {1 \over \epsilon^2} \, {\partial \over \partial \epsilon}
\left( \epsilon^2 \, {\cal D} \right) \eqno({\rm A}14)
$$
for the mean shear acceleration rate, in agreement with equation~(A2).

\vfil
\eject

\centerline{REFERENCES}

{
\refs

Beall, J. H., \& Bednarek, W. 1999, ApJ, submitted

Becker, P. A., \& Kafatos, M. 1995, ApJ, 453, 83

Becker, P. A., Kafatos, M., \& Maisack, M. 1994, ApJ Supplement,
90, 949

Begelman, M. C. 1998, ApJ, 493, 291

Begelman, M. C., \& Chiueh, T. 1988, ApJ, 332, 872

Blandford, R. D., \& Begelman, M. C. 1999, MNRAS, submitted

Blandford, R. D., \& Konigl, A. 1979, ApJ, 232, 34

Blandford, R. D., \& Payne, D. G. 1982, MNRAS, 199, 903

Bondi, H. 1952, MNRAS, 112, 195

Brandenburg, A., Nordlund, A., Stein, R. F., Torkelsson, U. 
1995, ApJ, 446, 741

Chakrabarti, S. K. 1985, ApJ, 288, 1

Contopoulos, J. 1995, ApJ, 446, 67

Contopoulos, J., \& Kazanas, D. 1995, ApJ, 441, 521

Contopoulos, J., \& Lovelace, R. V. E. 1994, ApJ, 429, 139

Daly, R. A., \& Marscher, A. P. 1988, ApJ, 334, 539

Dar, A., \& Laor, A. 1997, ApJ 478, L5

Dermer, C. D., Miller, J. A., \& Li, H. 1996, ApJ 456, 106

Dermer, C. D \& Schlickeiser, R. 1992, Science, 257, 1642

Earl, J. A., Jokipii, J. R., \& Morfill, G. 1988, ApJ, 331, L91

Eilek, J. A., \& Kafatos, M. 1983, ApJ, 271, 804

Frank, J., King, A. R., \& Raine, D. J. 1992, Accretion Power in
Astrophysics (Cambridge: Cambridge University Press)

Hartman, R. C., et al. 1992, ApJ, 385, L1

Hawley, J. F., Gammie, C. F., Balbus, S. A. 1995, ApJ, 440, 742

Katz, J. I. 1991, ApJ, 367, 407

Kazanas, D., \& Ellison, D. C. 1986, ApJ, 304, 178

Landau, L. D., \& Lifshitz, E. M. 1987, Fluid Mechanics
(New York: Pergamon Press)

Li, Z.-Y., Chiueh, T., \& Begelman, M. C. 1992, ApJ, 394, 459

Lynden-Bell, D. 1996, MNRAS, 279, 389

Mannheim, K. 1993, A \& A, 269, 67

Mastichiadis, A., \& Kazanas, D. 1993, BAAS, 182, 12.07

Matsumoto, R., \& Tajima, T. 1995, ApJ, 445, 767

M\'esz\'aros, P., \& Rees, M. J. 1992, MNRAS, 257, 29

Mihalas, D., \& Mihalas, B. W. 1984, Foundations of Radiation Hydrodynamics
(New York: Oxford University Press)

Narayan, R., \& Yi, I. 1995, ApJ, 444, 231

Paczynski, B., \& Wiita, P. J. 1980, A \& A, 88, 23

Phinney, E. S. 1982, MNRAS, 198, 1109

Protheroe, R. J., \& Kazanas, D. 1983, ApJ, 265, 620

Rees, M. J., Begelman, M. C., Blandford, R. D., \& Phinney,
E. S. 1982, Nature, 295, 17

Reif, F. 1965, Fundamentals of Statistical and Thermal Physics
(New York: McGraw-Hill)

Romanova, M. M., Ustyugova, G. V., Koldoba, A. V., Chechetkin,
V. M., \& Lovelace, R. V. E. 1998, ApJ, 500, 703

Rybicki, G. B., \& Lightman, A. P. 1979, Radiative Processes in
Astrophysics (New York: Wiley)

Shapiro, S. L., Lightman, A. L., \& Eardley, D. M. 1976, ApJ,
204, 187

Schmidt, G. 1966, Physics of High Temperature Plasmas (New York:
Academic)

Subramanian, P. 1997, PhD Thesis, George Mason University

Subramanian, P., Becker, P. A.,  \& Kafatos, M. 1996, ApJ,
469, 784 (Paper 1)

Ustyugova, G. V., Koldoba, A. V., Romanova, M. M., Chechetkin,
V. M., \& Lovelace, R. V. E. 1995, ApJ, 439, L39

Webb, G. M., Jokipii, J. R., \& Morfill, G. E. 1994, ApJ, 424,
158

Wehrle, A. E. et al. 1998, ApJ, 497, 178

Weinberg, S. 1972, Gravitation and Cosmology: Principles and
Application of the General Theory of Relativity (New York: Wiley)

Vermeulen, R. C., \& Cohen, M. H. 1994, ApJ, 430, 467

von Montigny, C. et al. 1995, ApJ, 440, 525

}

\vfil
\eject

\phantom{XXXXXXXXXXXX}
\phantom{XXXXXXXXXXXX}
\phantom{XXXXXXXXXXXX}
\phantom{XXXXXXXXXXXX}
\phantom{XXXXXXXXXXXX}

\centerline{TABLE 1}
\centerline{Global Model Parameters}
\medskip
\hskip 0.8truein
\vbox{
\hsize=5.0truein
\hrule
\smallskip
\hrule
\+\ &\phantom{STUFF} &\phantom{STUFF}    &\phantom{STUFF}
&\phantom{STUFFF}  &\phantom{STUFFFF}  &\phantom{STUF}
&\phantom{STUFF} \cr
\vskip-0.2truein
\+\ &Model   &~$\theta$    &~$\xi_0$    &$\Rin / R_g$
&$\Rout / R_g$   &$\langle \Gamma_\infty \rangle$  &$\Min/\Mout$ \cr

\medskip
\hrule
\medskip

\+\ &~~~1   &0.5$^\circ$    &0.05    &~~4.47   &~~7.37   &6.076  &~~~0.835 \cr

\+\ &~~~2   &1.0$^\circ$    &0.05    &~~4.95   &~~8.28   &5.765  &~~~0.826 \cr

\+\ &~~~3   &0.5$^\circ$    &0.1    &~~3.55   &~~5.25   &4.244  &~~~0.764  \cr

\+\ &~~~4   &1.0$^\circ$    &0.1    &~~3.91   &~~5.97   &3.929  &~~~0.745  \cr

\medskip
\hrule
}

\vfil
\eject

\phantom{XXXXXXXXXXXX}
\phantom{XXXXXXXXXXXX}
\phantom{XXXXXXXXXXXX}
\phantom{XXXXXXXXXXXX}
\phantom{XXXXXXXXXXXX}

\centerline{~~~~~~~~TABLE 2}
\centerline{~~~~~~~~Model Parameters for $R = \Rout$}
\medskip
\hskip 0.8truein
\vbox{
\hsize=5.0truein
\hrule
\smallskip
\hrule
\+\ &\phantom{STUFF} &\phantom{STUFF}    &\phantom{STUFF}
&\phantom{STUFFF}  &\phantom{STUFFFF}  &\phantom{STUF}
&\phantom{STUFF} \cr
\vskip-0.2truein
\+\ &Model   &~$z_1/R_g$    &~~$v_1/c$    &~~~~$s_1/c$
&~~~$z_c/R_g$    &~~$v_c/c$  &~~~~$s_c/c$ &~~~~$\Gamma_{\infty}$ \cr

\medskip
\hrule
\medskip

\+\ &~~~1   &~~20.9    &~~0.48    &~~~~0.89   &~~~~22.8   &~~0.48  &~~~~0.89 
&~~~4.24 \cr

\+\ &~~~2   &~~26.9    &~~0.53    &~~~~0.86   &~~~~16.5   &~~0.48  &~~~~0.88 
&~~~4.16\cr

\+\ &~~~3   &~~9.2    &~~0.33    &~~~~0.68   &~~~~22.5   &~~0.43  &~~~~0.64
&~~~2.96 \cr

\+\ &~~~4   &~~12.6    &~~0.39    &~~~~0.66   &~~~~17.6   &~~0.43  &~~~~0.65
&~~~2.88 \cr

\medskip
\hrule
}

\vfil
\eject

\centerline{\bf FIGURE CAPTIONS}

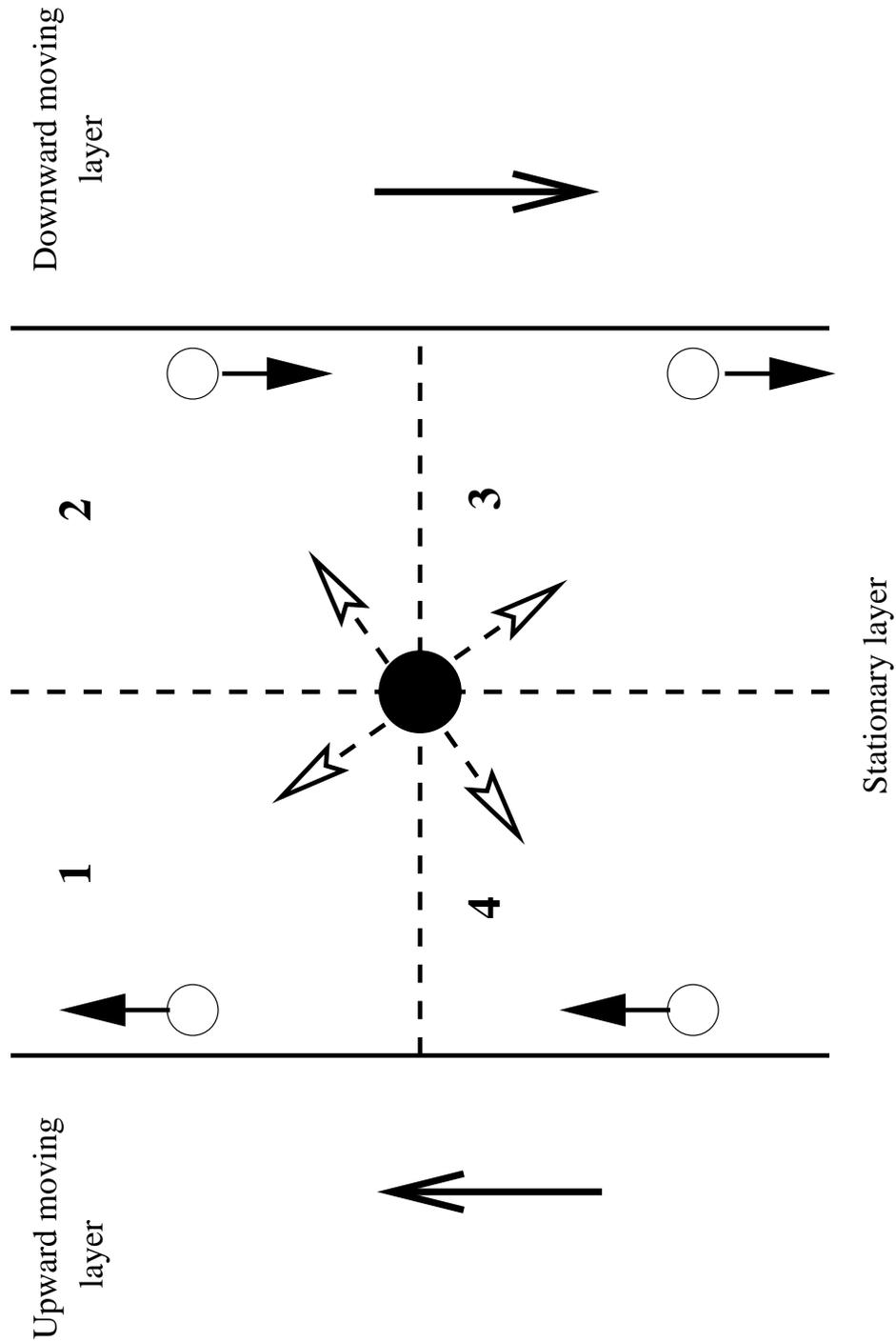
Fig. 1. -- Schematic depiction of second-order Fermi acceleration
resulting from an average collision between a proton originating
at the (stationary) origin and a scattering center (cold magnetic kink)
located in one of the four quadrants. The solid circle at the center
represents the incident proton, and the open circles in the four
quadrants represent the scattering centers, which move with the velocity
of the shear flow.

\bigskip

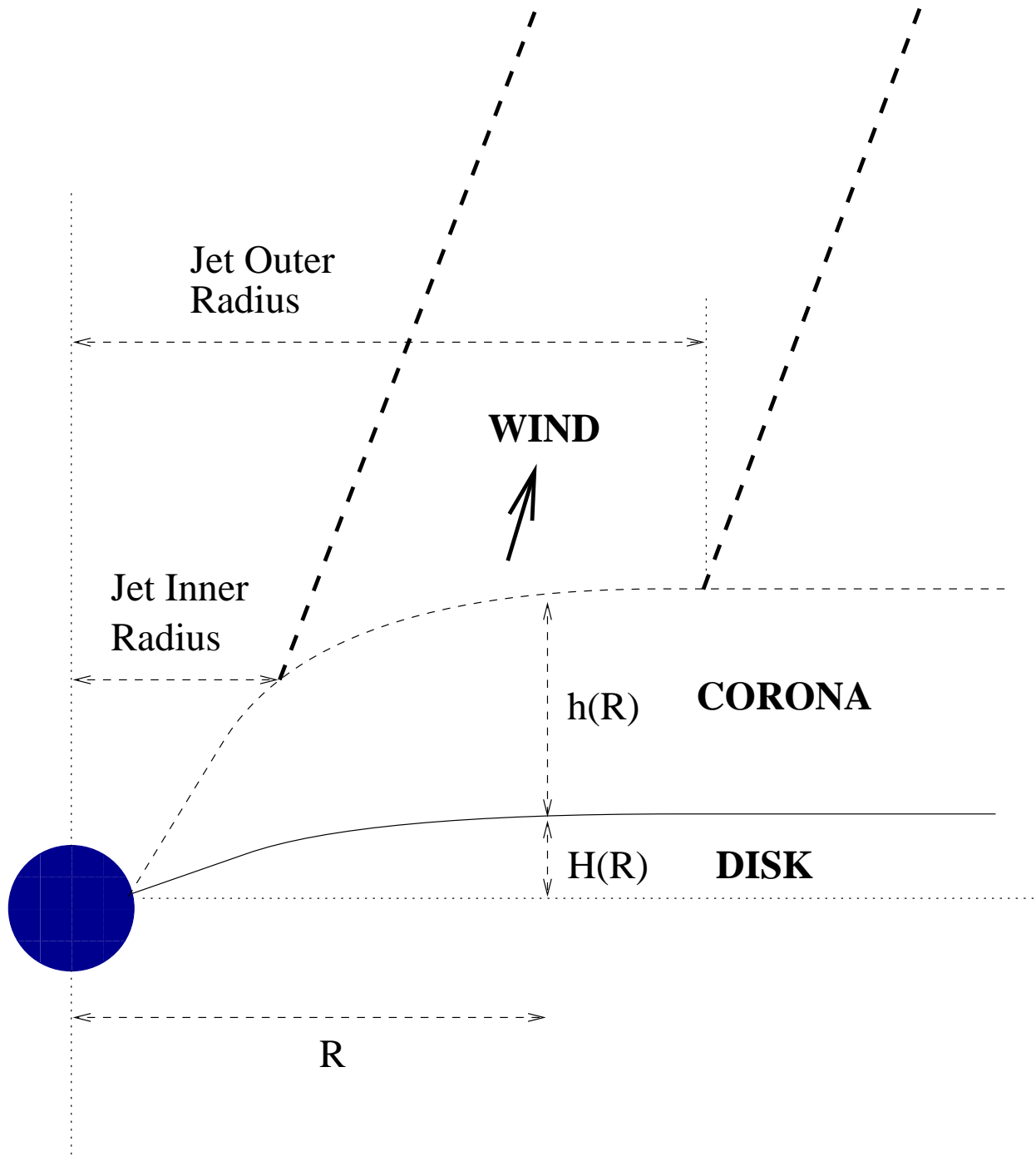
Fig. 2. -- Schematic diagram of the disk/corona/wind geometry.

\bigskip

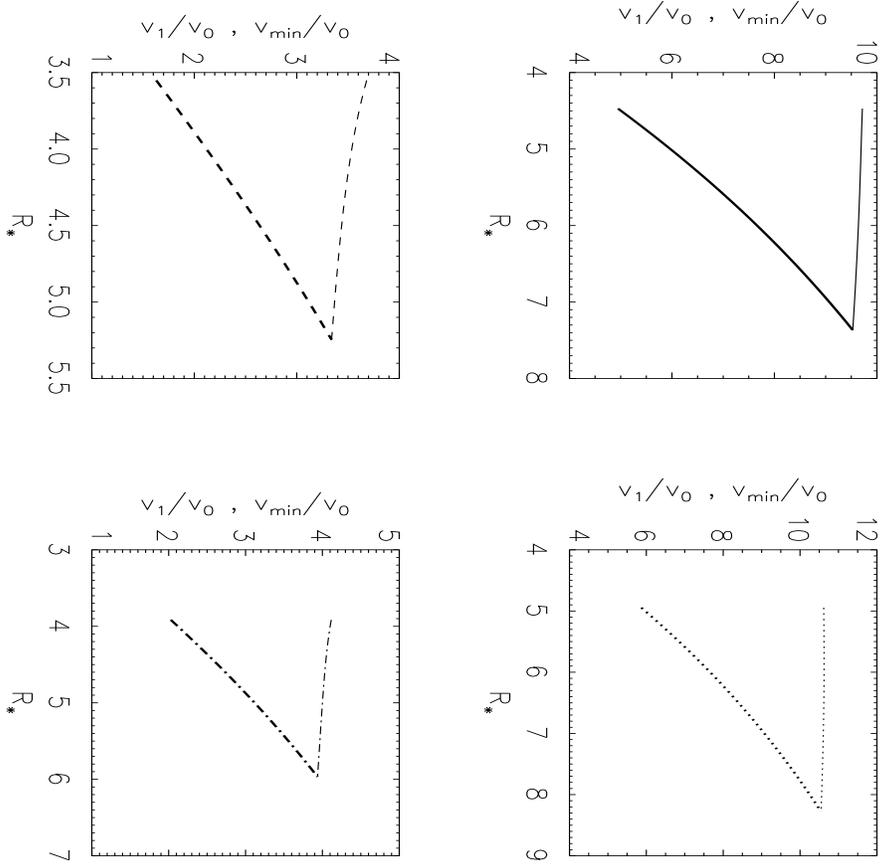
Fig. 3. -- Comparison of $v_{1}$/$v_{0}$ and $v_{\rm min}$/$v_{0}$
as functions of the dimensionless cylindrical starting radius in
the corona, $R_*$, for each of the computed models. The point
of intersection of these two curves defines the outer radius
(see eq.~[6.6]). The thin lines represent $v_{1}$/$v_{0}$ and
the heavy lines represent $v_{\rm min}$/$v_{0}$. Throughout
the results presented here, we will use a uniform set of line
styles for Model 1 ({\it solid lines}), Model 2 ({\it dotted lines}),
Model 3 ({\it dashed lines}), and Model 4 ({\it dash-dot lines}).

\bigskip

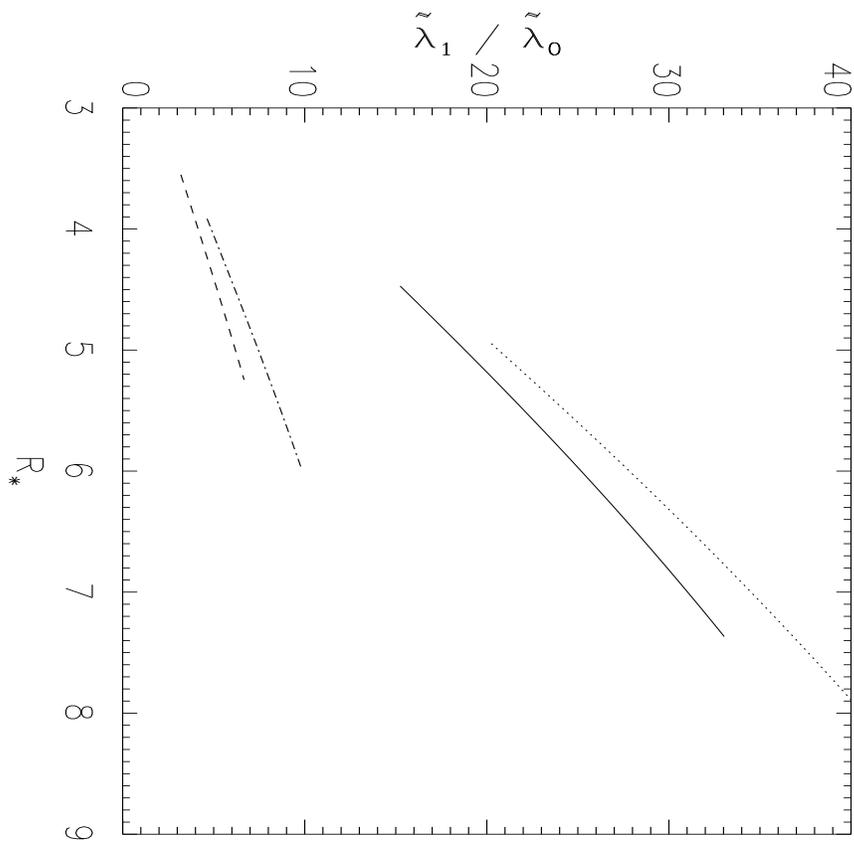
Fig. 4. -- Ratio of magnetic coherence lengths in the corona and disk,
$\tilde \lambda_{1}$/$\tilde \lambda_{0}$,
plotted as a function of the dimensionless starting radius
$R_*$ for each of the computed models. In each case the field
lines are less tangled in the corona than they are in the disk.

\bigskip

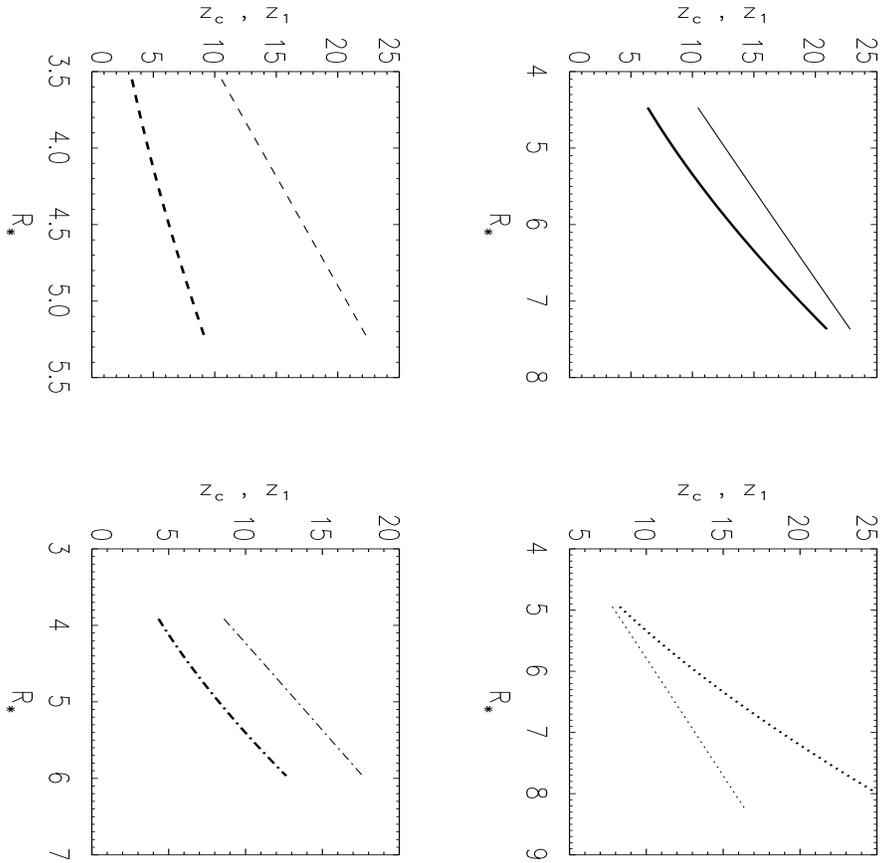
Fig. 5. -- Comparison of the height of the critical surface
$z_{c}$ and the starting height for the wind $z_{1}$ as functions
of the dimensionless starting radius $R_*$ for each of the
models. The thin lines represent $z_{c}/R_g$ and the
heavy lines represent $z_{1}/R_g$. Note that in the case of Model 2
the wind is supersonic at the base. In all of the other cases, the
wind is subsonic at the base.

\bigskip

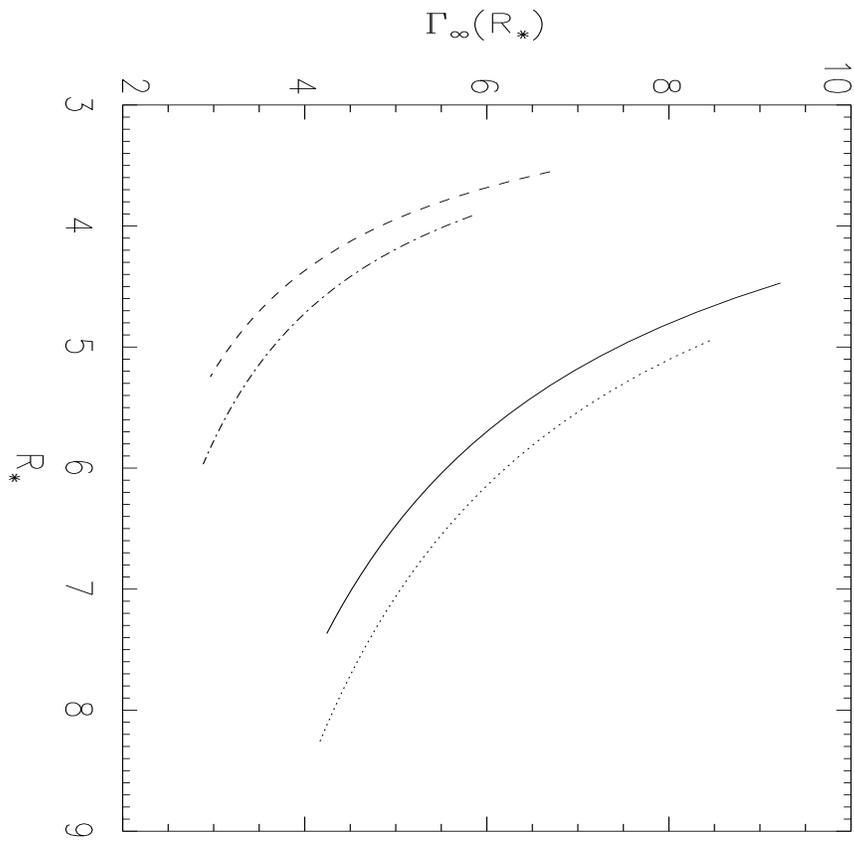
Fig. 6. -- Asymptotic Lorentz factor $\Gamma_{\infty}$ plotted as
a function of the starting radius $R_*$ using eq.~(5.21). 
For all of the models, the asymptotic Lorentz factor tends to
decrease with increasing radius due to the diminishing strength of the
Keplerian shear.

\bigskip

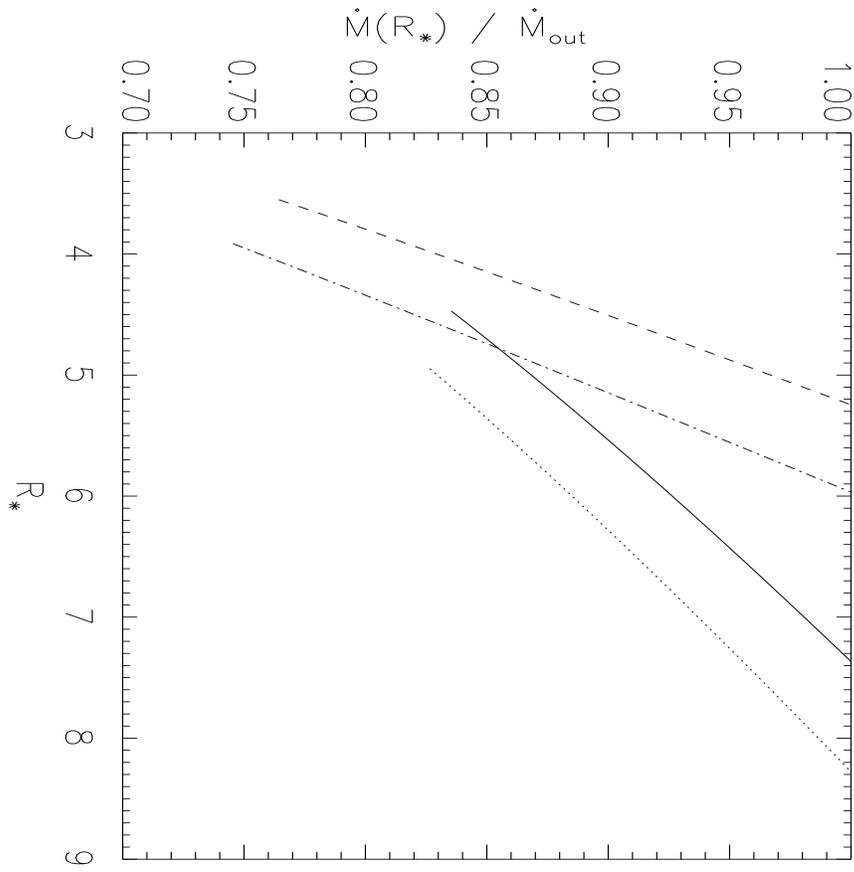
Fig. 7. -- Radial variation of the disk accretion rate $\dot{M}$ as
a function of $R_{*}$ plotted using eq.~(4.23) for each of the models.
The accretion rate decreases with increasing radius in response to the
mass lost into the jet and the counterjet.

\bigskip

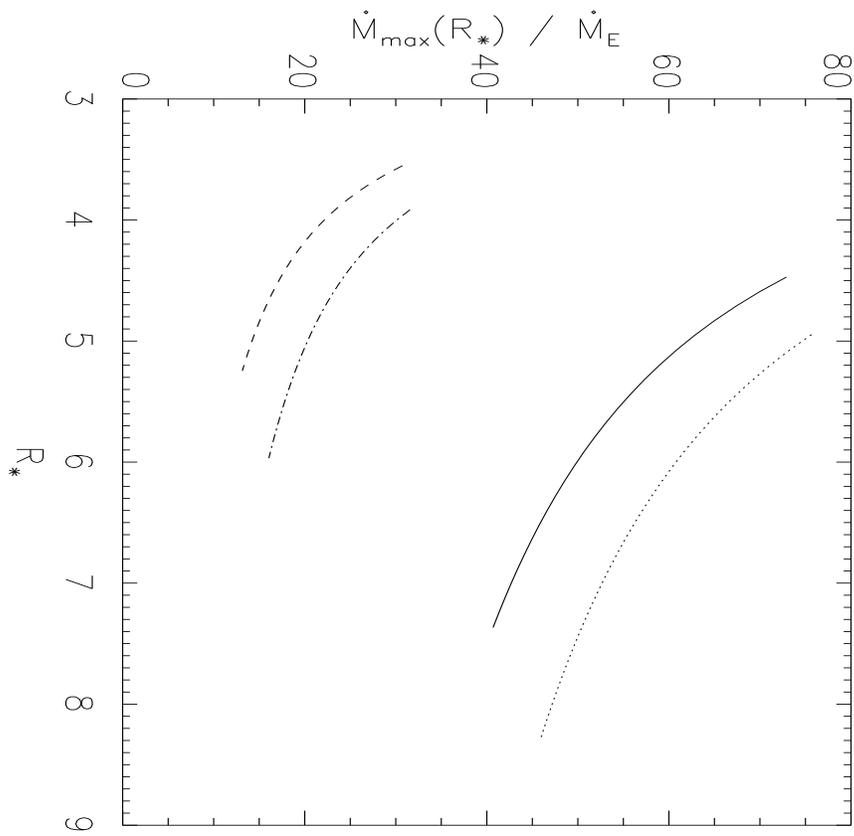
Fig. 8. -- Upper limit on the accretion rate $\dot{M}_{\rm max}
/\dot{M}_{\rm E}$ plotted as a function of $R_{*}$ using eq.~(6.10)
for each of the models. For accretion rates exceeding $\dot{M}_{\rm max}$,
losses due to strong proton-proton collisions in the corona overwhelm the
shear acceleration.

\bigskip

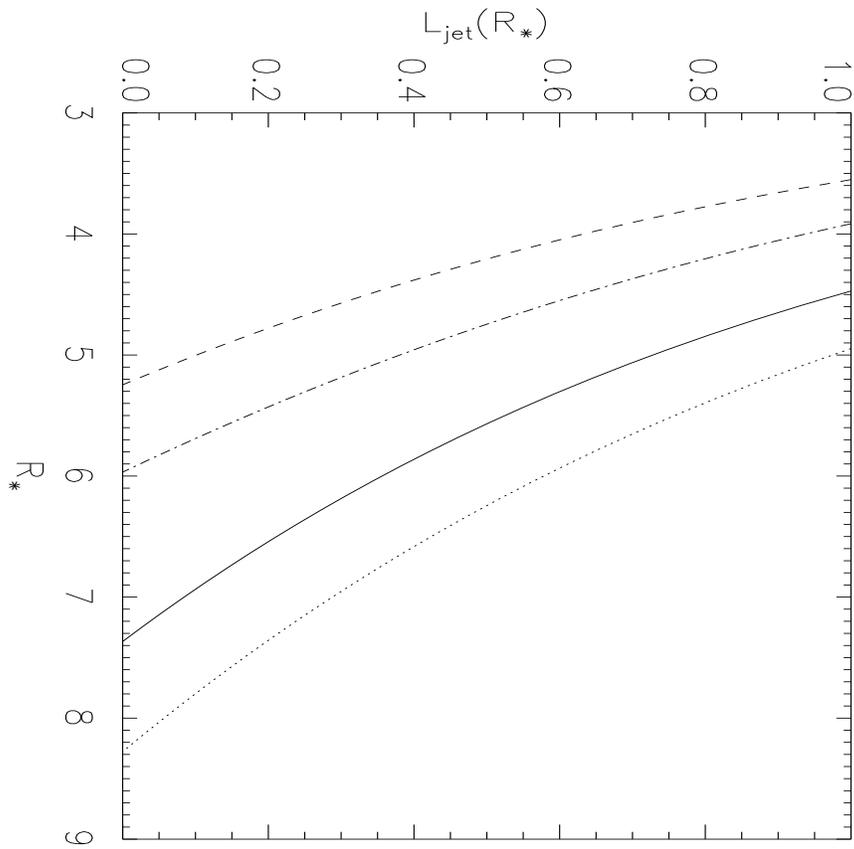
Fig. 9. -- Asymptotic kinetic luminosity of the jet $L_{\rm jet}$
(eq.~[6.16]) plotted in units of the maximum accretion luminosity
(1/2)$\dot{M}_{\rm out} \, c^{2}$ as a function of $R_{*}$. The inner
radius of the jet, $R_{\rm in}$, is defined as the radius at which
this quantity reaches unity (see Table 1).

\bigskip

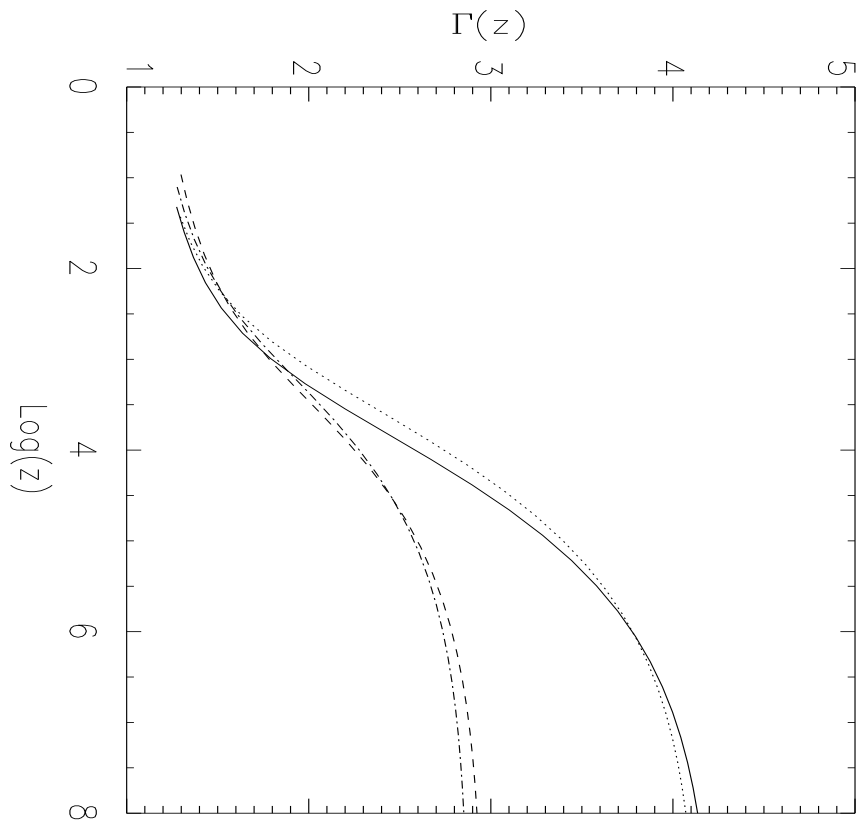
Fig. 10. -- Lorentz factor $\Gamma$ (eq.~[5.7]) plotted as a
function of the height $z/R_g$ measured from the midplane of the
disk for starting radius $R = R_{\rm out}$. Note that the flow
continues to accelerate at large distances from the black hole.

\bigskip

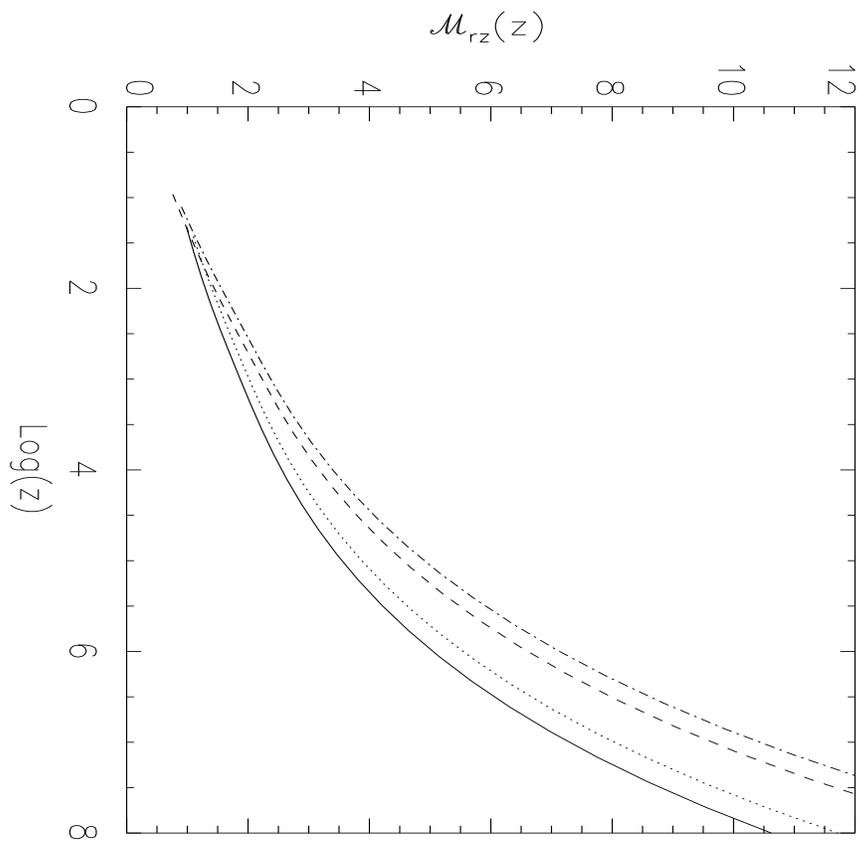
Fig. 11. -- Mach number in the $(r,z)$ plane, ${\cal M}_{r \, z}$,
plotted as a function of the height $z/R_g$ above the midplane of
the disk using eq.~(7.3) for starting radius $R = R_{\rm out}$.

\bigskip

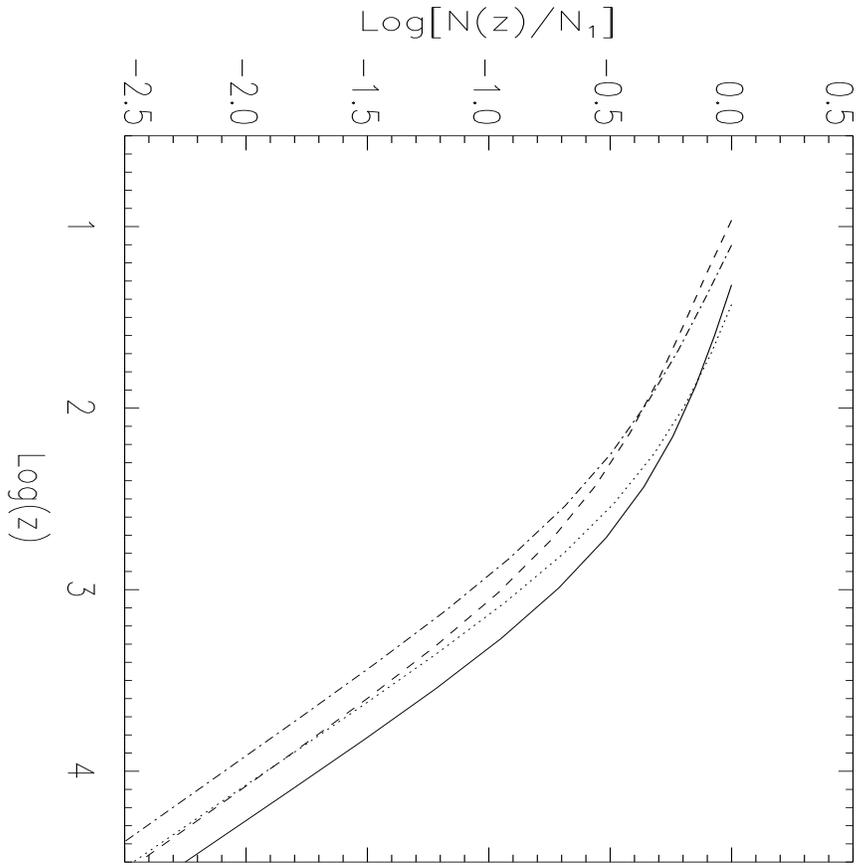
Fig. 12. -- Variation of the number density in the wind $N$ plotted as
a function of the height $z/R_g$ above the disk midplane for starting
radius $R = R_{\rm out}$.

\bigskip

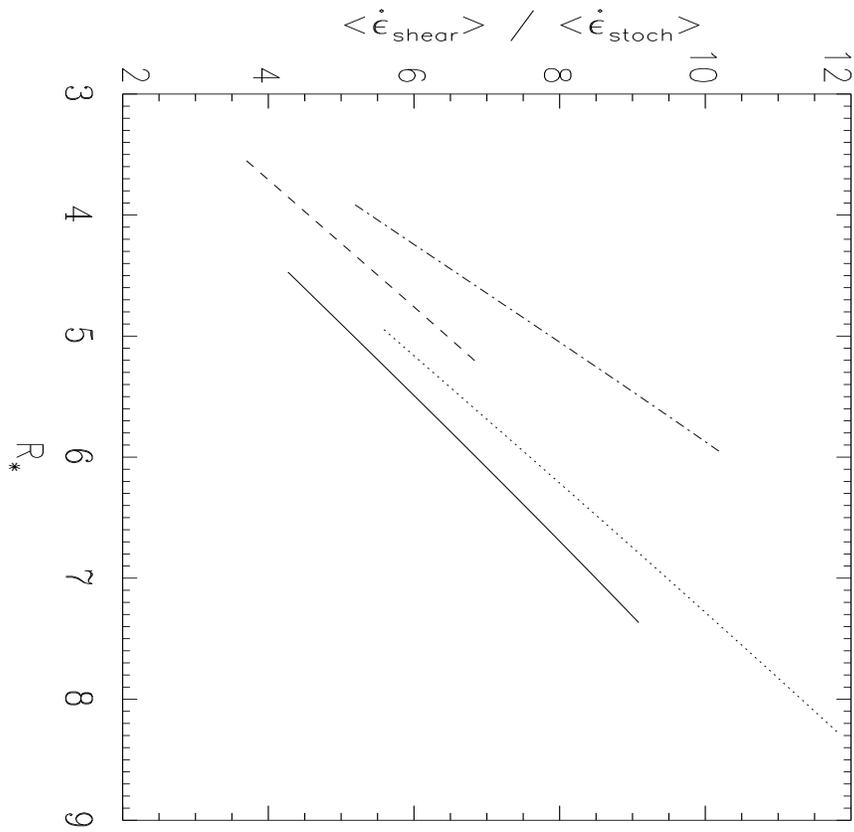
Fig. 13. -- Ratio of the shear acceleration rate $\langle
\dot{\epsilon}_{\rm shear} \rangle$ to the stochastic acceleration
rate $\langle \dot{\epsilon}_{\rm stoch} \rangle$ plotted as a
function of $R_{*}$, evaluated using eq.~(8.8). Note that
$\langle \dot{\epsilon}_{\rm shear} \rangle$ greatly
exceeds $\langle \dot{\epsilon}_{\rm stoch} \rangle$ at all radii.

\bye